%% file: main.tex
\documentclass[unnumsec,webpdf,contemporary,large]{main}%

\usepackage{color,soul}
\usepackage[british]{babel}
\usepackage{booktabs}

\newcommand{\xhdr}[1]{\vspace{1.7mm}\noindent{{\bf #1.}}}
\DeclareCaptionType{InfoBox}
\graphicspath{{Fig/}}

\theoremstyle{thmstyleone}

\theoremstyle{thmstyletwo}%
\theoremstyle{thmstylethree}%

\begin{document}

\journaltitle{Bioinformatics Advances}
\DOI{DOI HERE}
\copyrightyear{2024}
\pubyear{2024}
\access{Advance Access Publication Date: Day Month Year}
\appnotes{Perspective}

\firstpage{1}

\title[Current and future directions in network biology]{Current and future directions in network biology}
\renewcommand\rightmark{Current and future directions in network biology}
\renewcommand\leftmark{M. Zitnik \textit{et. al.}}

\author[1,$\dagger$]{Marinka Zitnik}
\author[1,$\dagger$]{Michelle M.~Li}
\author[2,3,4,$\dagger$]{Aydin Wells}
\author[5,\#]{Kimberly Glass}
\author[5,6,7,\#]{Deisy Morselli Gysi}
\author[8,\#]{Arjun Krishnan}
\author[9,\#]{T. M. Murali}
\author[10,\#]{Predrag Radivojac}
\author[\text{11,12},\#]{Sushmita Roy}
\author[13]{Ana\"{i}s Baudot}
\author[14,15]{Serdar Bozdag}
\author[2]{Danny Z. Chen}
\author[16]{Lenore Cowen}
\author[16]{Kapil Devkota}
\author[\text{11,17}]{Anthony Gitter}
\author[18]{Sara Gosline}
\author[2]{Pengfei Gu}
\author[19]{Pietro H. Guzzi}
\author[20]{Heng Huang}
\author[2]{Meng Jiang}
\author[\text{14,21}]{Ziynet Nesibe Kesimoglu}
\author[\text{22}]{Mehmet Koyuturk}
\author[23]{Jian Ma}
\author[24]{Alexander R. Pico}
\author[25,26,27]{Nata\v{s}a Pr\v{z}ulj}
\author[21]{Teresa M. Przytycka}
\author[28]{Benjamin J. Raphael}
\author[29]{Anna Ritz}
\author[30]{Roded Sharan}
\author[31]{Yang Shen}
\author[28,32]{Mona Singh}
\author[16]{Donna K. Slonim}
\author[\text{33}]{Hanghang Tong}
\author[34]{Xinan Holly Yang}
\author[31,35]{Byung-Jun Yoon}
\author[36]{Haiyuan Yu}
\author[2,3,4,*]{Tijana Milenkovi\'{c}}

\address[1]{\orgdiv{Department of Biomedical Informatics}, \orgname{Harvard University}}

\address[2]{\orgdiv{Department of Computer Science and Engineering}, \orgname{University of Notre Dame}}

\address[3]{\orgdiv{Lucy Family Institute for Data and Society}, \orgname{University of Notre Dame}}

\address[4]{\orgdiv{Eck Institute for Global Health}, \orgname{University of Notre Dame}}

\address[5]{\orgdiv{Channing Division of Network Medicine}, \orgname{Brigham and Women’s Hospital, Harvard Medical School}}

\address[6]{\orgdiv{Department of Statistics}, \orgname{Federal University of Paraná}}

\address[7]{\orgdiv{Department of Physics}, \orgname{Northeastern University}}

\address[8]{\orgdiv{Department of Biomedical Informatics}, \orgname{University of Colorado}}

\address[9]{\orgdiv{Department of Computer Science}, \orgname{Virginia Tech}}

\address[10]{\orgdiv{Khoury College of Computer Sciences}, \orgname{Northeastern University}}

\address[\text{11}]{\orgdiv{Department of Biostatistics and Medical Informatics}, \orgname{University of Wisconsin-Madison}}

\address[12]{\orgname{Wisconsin Institute for Discovery}}

\address[13]{\orgname{Aix Marseille Université, INSERM, MMG}, \orgaddress{Marseille, France}}

\address[14]{\orgdiv{Department of Computer Science and Engineering}, \orgname{University of North Texas}}

\address[15]{\orgdiv{Department of Mathematics}, \orgname{University of North Texas}}

\address[16]{\orgdiv{Department of Computer Science}, \orgname{Tufts University}}

\address[17]{\orgname{Morgridge Institute for Research}}

\address[18]{\orgname{Pacific Northwest National Laboratory}}

\address[19]{\orgdiv{Department of Medical and Surgical Sciences}, \orgname{University Magna Graecia of Catanzaro}}

\address[20]{\orgdiv{Department of Computer Science}, \orgname{University of Maryland College Park}}

\address[21]{\orgdiv{National Center of Biotechnology Information}, \orgname{National Library of Medicine, National Institutes of Health}}

\address[\text{22}]{\orgdiv{Department of Computer and Data Sciences}, \orgname{Case Western Reserve University}}

\address[23]{\orgdiv{Computational Biology Department, School of Computer Science}, \orgname{Carnegie Mellon University}}

\address[24]{\orgdiv{Institute of Data Science and Biotechnology}, \orgname{Gladstone Institutes}}

\address[25]{\orgdiv{Department of Computer Science}, \orgname{University College London}}

\address[26]{\orgdiv{ICREA}, \orgname{Catalan Institution for Research and Advanced Studies}}

\address[27]{\orgname{Barcelona Supercomputing Center (BSC)}}

\address[28]{\orgdiv{Department of Computer Science}, \orgname{Princeton University}}

\address[29]{\orgdiv{Department of Biology}, \orgname{Reed College}}

\address[30]{\orgdiv{School of Computer Science}, \orgname{Tel Aviv University}}

\address[31]{\orgdiv{Department of Electrical and Computer Engineering}, \orgname{Texas A{\&}M University}}

\address[32]{\orgdiv{Lewis-Sigler Institute for Integrative Genomics}, \orgname{Princeton University}}

\address[\text{33}]{\orgdiv{Department of Computer Science}, \orgname{University of Illinois Urbana-Champaign}}

\address[34]{\orgdiv{Department of Pediatrics}, \orgname{University of Chicago}}

\address[35]{\orgdiv{Computational Science Initiative}, \orgname{Brookhaven National Laboratory}}

\address[36]{\orgdiv{Department of Computational Biology}, \orgname{Weill Institute for Cell and Molecular Biology, Cornell University}}

\corresp[$\dagger$]{Co-first authors. }

\corresp[$\#$]{Co-second authors: coordinators for Sections 2-7, listed in their alphabetical order by last name. }

\corresp[$\ast$]{Corresponding author  (email: \href{email:tmilenko@nd.edu}{tmilenko@nd.edu}).} 

\corresp[]{All other authors are listed in their alphabetical order by last name. \\ Co-authorships on this paper are the result of the co-authors participating in the same workshop and not of scientific collaboration of any sort. As such, the co-authorships do not constitute any conflict of interest.}

\abstract{Network biology is an interdisciplinary field bridging computational and biological sciences that has proved pivotal in advancing the understanding of cellular functions and diseases across biological systems and scales. Although the field has been around for two decades, it remains nascent. It has witnessed rapid evolution, accompanied by emerging challenges. These challenges stem from various factors, notably the growing complexity and volume of data together with the increased diversity of data types describing different tiers of biological organization.
We discuss prevailing research directions in network biology and highlight areas of inference and comparison of biological networks, multimodal data integration and heterogeneous networks, higher-order network analysis, machine learning on networks, and network-based personalized medicine. Following the overview of recent breakthroughs across these five areas, we offer a perspective on the future directions of network biology. Additionally, we offer insights into scientific communities, educational initiatives, and the importance of fostering diversity within the field. This paper establishes a roadmap for an immediate and long-term vision for network biology.}


\maketitle

\section*{Acronyms used in this paper}
3D: 3-dimensional\\
BKG: biomedical knowledge graph\\ 
CAFA: Critical Assessment of protein Function Annotation\\ 
CASP: Critical Assessment of protein Structure Prediction\\ 
CondBKG: condition-aware biomedical knowledge graph\\ 
DREAM: Dialogue on Reverse Engineering Assessment and Methods\\ 
EDI: equity, diversity, and inclusion\\ 
GDV: graphlet degree vector\\ 
GO: Gene Ontology\\
GNN: graph neural network\\ 
ISCB: International Society for Computational Biology\\ 
LLM: large language model\\
PPI: protein-protein interaction\\ 
TF: transcription factor\\ 
TGF$\beta$: transforming growth factor-beta

\section{1. Introduction}\label{sect:introduction}
\input{010.intro_rtf}

\section{2. Inference and comparison of biological networks}\label{sect:network_inference_comparison}
\input{020.inference_comparison_rtf}

\section{3. Multimodal data integration and heterogeneous networks}\label{sect:multimodal_networks_integration}
\input{030.multimodal_rtf}

\section{4. Higher-order network analysis}\label{sect:higher_order_network_analysis}
\input{040.higher_order_rtf}

\section{5. Machine learning on networks}\label{sect:networkml-intro}
\input{050.machine_learning_rtf}

\section{6. Network-based personalized medicine}\label{sect:personalized_medicine}
\input{060.personalized_medicine_rtf}

\section{7. Research discussion and future outlook}\label{sect:research_discussion_future_directions}
\input{070.discussion_rtf}

\section{8. Additional discussion on scientific communities, education, and diversity}\label{sect:additional_discussion}
\input{080.community_rtf}

\section{Software and data availability}
Not applicable.

\section{Conflict of interest}
In Section \ref{sect:additional_discussion}, we rely on ISCB's diversity statistics. These statistics are publicly available, and so there is no conflict of interest. Yet, to remedy any potential perceived conflict of interest, we declare that Predrag Radivojac is the President of ISCB and currently serves on the Board of Directors of ISCB. In addition, Tijana Milenkovi\'{c} currently serves on the ISCB Board of Directors and the ISCB EDI Committee. The remaining authors have no conflicts of interest to declare.

\section{Acknowledgements}
This work has been initialized at the Workshop on Future Directions in Network Biology held at the University of Notre Dame during June 12-14, 2022. The workshop was supported by the U.S. National Science Foundation [grant number CCF-1941447]. This targeted meeting brought together 39 active researchers in various aspects of network biology to present and discuss a short- and long-term vision for computational research in this field. 31 of the workshop participants attended the meeting in person. Due to difficulties with international travel related to the COVID-19 pandemic, all in-person workshop participants were from institutions in the United States. To draw on a combination of distinct ideas and experiences, when inviting participants, an effort was made to balance diversity among the attendees along multiple axes, including seniority (full, associate, or assistant professors, postdocs, and PhD students), affiliation (representation from academia, industry, and government), and gender (42\% of the in-person participants were female). 
The workshop participants presented their views of important research directions, open problems, and challenges that would propel computational and algorithmic advances in network biology. Presentation slides for the scientific sessions at the workshop are linked to the workshop website\footnote{\url{https://www3.nd.edu/~tmilenko/NetworkBiologyWorkshop/}}, and videos of the presentations are publicly available on YouTube\footnote{\url{https://www.youtube.com/playlist?list=PLy8BJXti_TvYaL7frFJz2mf38e8o0NaFN}}.

Thanks to Siyu Yang, a Ph.D. student in the Department of Computer Science and Engineering at the University of Notre Dame, for carrying out the literature search on network-of-networks analysis. 

Pacific Northwest National Laboratory is operated by Battelle for the U.S. Department of Energy under Contract No. DE-AC05 to 76RLO 1830. 

The work of Teresa M. Przytycka was supported by the Intramural Research Program of the National Library of Medicine, National Institutes of Health [grant number LM200887-16].

\bibliographystyle{abbrv}
\bibliography{Bib/Superbib,Bib/extra}

\end{document}

%% file: 010.intro_rtf.tex
A network (or graph) is comprised of a set of nodes (or vertices) that are connected by a set of edges (or links); see InfoBox~\ref{box:terminology}. Networks allow us to study the properties of a complex system that emerge from interactions between its individual components. Networks have been a powerful way to represent a variety of real-world phenomena, including technological, information, transportation, social, financial, software, ecological, chemical, and biological systems \cite{Barabasi2016,newman2018networks}. Our focus is on biological networks, which offer the understanding of complex functions at the levels of genes, proteins, cells, tissues, organs, etc., by representing a given biological system as an interconnected entity rather than a collection of individual components. In a biological network, nodes {typically} represent biomolecules (e.g., amino acid residues within a protein, proteins within a cell, or cells within a tissue), and edges {typically} indicate interactions between the biomolecules (e.g., physical, functional, or chemical).  {While the main focus of our paper is on such biological networks that model relationships between biomolecules, i.e., on molecular/cellular networks, our paper touches on other types of biological networks, such as biomedical knowledge graphs, ontologies, patient similarity networks modeling e.g., electronic health record data, brain networks constructed from medical imaging data, and even social and contact networks relevant for spread of disease. We acknowledge that other types of biological networks exist that are not the  focus of our paper and that we thus do not cover, such as ecological ones.}

\begin{figure*}[t!]
  \centering
    \includegraphics[width=\linewidth]{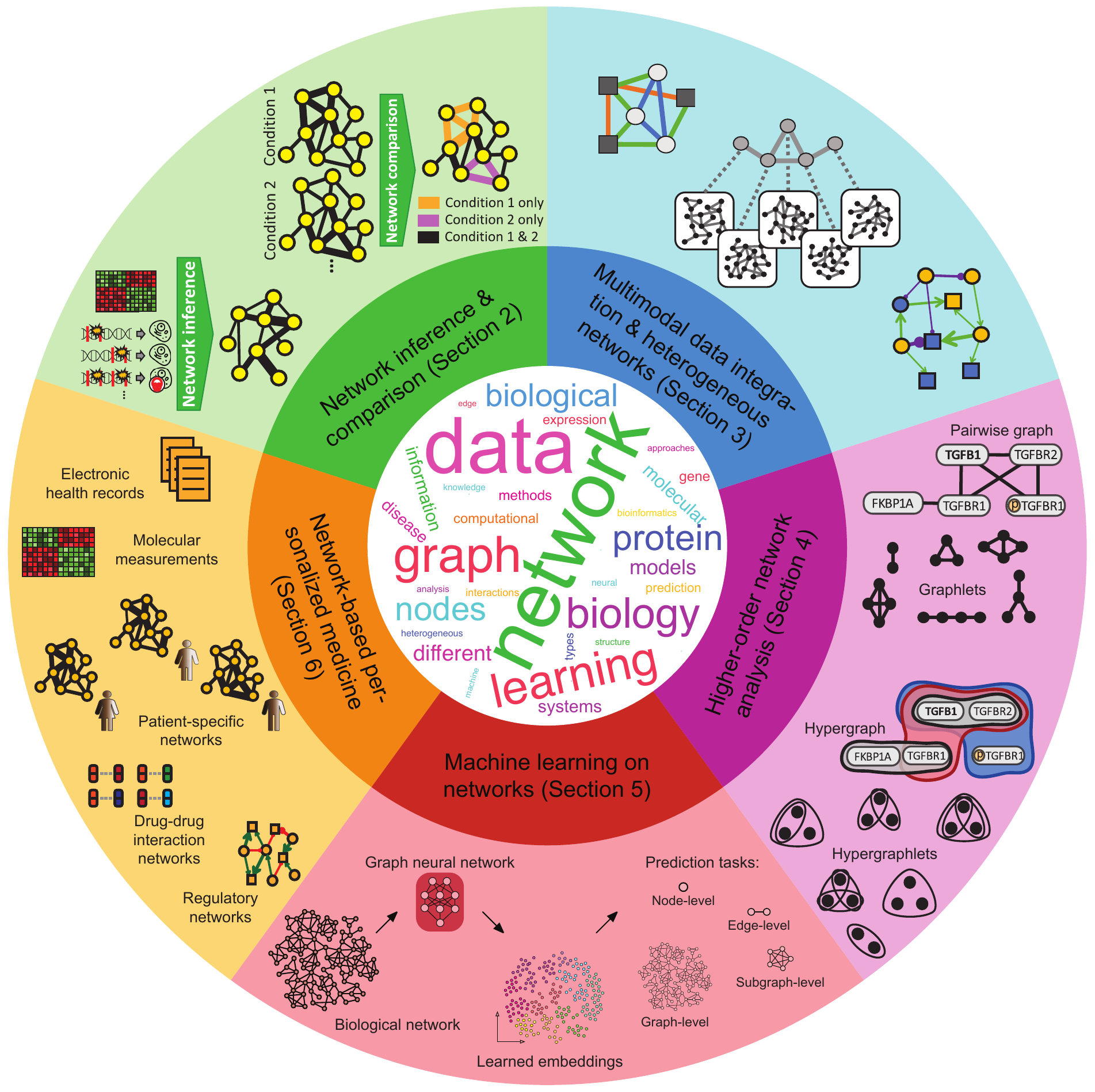}
  \caption{Overview of the network biology field and five research topics discussed in this paper. The word cloud in the center, generated using \url{WordClouds.com}, contains the top 30 most representative words from this paper. Note that each word's rank is based on the sum of the weights of the core word (e.g., learn) and its derived words (e.g., learns, learning, learned).}
  \label{fig:overview}
\end{figure*}

Network biology (Fig. \ref{fig:overview}) is an interdisciplinary field spanning computational (e.g., algorithms, graph theory, network science, data mining, and machine learning) and biological sciences. While the field has existed for nearly two decades, it has undergone numerous rapid changes and new computational challenges have arisen. This is caused by many factors, including increasing data complexity, such as multiple types of data becoming available at different levels (or scales) of biological organization, as well as growing data size. Ironically, despite the massive increase in available data, the data remain incomplete and noisy.  This means that the research directions in the field also need to evolve. 

This paper discusses the current state as well as the future of the field. Its goal  is to identify pressing challenges with well-established as well as emerging topics in network biology, which are shown in Fig. \ref{fig:overview}: inference and comparison of biological networks~(Section~\ref{sect:network_inference_comparison}), multimodal data integration and heterogeneous networks~(Section~\ref{sect:multimodal_networks_integration}), higher-order network analysis~(Section~\ref{sect:higher_order_network_analysis}), machine learning on networks~(Section~\ref{sect:networkml-intro}), and network-based personalized medicine~(Section~\ref{sect:personalized_medicine}). We comment on why these topics have been strategically chosen for discussion in this paper.

{Noting again that a key focus of our paper is on molecular/cellular (i.e., -omics) data,} certain types of -omics data are explicitly captured as networks. That is, interactions between biomolecules are measured explicitly by biotechnological data collection platforms. A prominent example is protein-protein interaction (PPI) networks. In these networks, nodes are proteins and edges correspond to physical bindings between the proteins. In human and some model organisms, extensive high-throughput yeast two-hybrid and other experimental efforts have resulted in large sets of ``reference'' PPIs (such as HURI for human), along with substantial knowledge about protein binding specificities \cite{luck2020reference,stark2006biogrid}. 

Other types of -omics data are not captured as networks explicitly, but interactions between biomolecules can be inferred computationally, resulting in, e.g., association, correlation, regulatory, or knowledge graphs (InfoBox~\ref{box:terminology}). Section \ref{sect:network_inference_comparison} addresses several aspects of the task of inferring a homogeneous network, including a condition-specific network, typically from up to a couple of -omics data types/modes, along with a related topic of differential network analysis, which is one type of network comparison. Section \ref{sect:multimodal_networks_integration} addresses the task of inferring a heterogeneous network, typically from diverse -omics or other multimodal data types (InfoBox~\ref{box:terminology}), along with several other tasks related to multi-omics data integration, including network alignment, which is another type of network comparison. By a homogeneous network, we mean a network with a single node type and a single edge type, while by a heterogeneous network, we mean any non-homogeneous network (i.e., multiple node types or multiple edge types or both); see InfoBox~\ref{box:terminology} and Section \ref{sect:multimodal_networks_integration} for details. 

Given (explicitly captured or inferred) network data, the next step is to analyze the data.  While Sections \ref{sect:network_inference_comparison} and \ref{sect:multimodal_networks_integration} already address network analysis from the perspective of network comparison and several other tasks, Sections \ref{sect:higher_order_network_analysis} and \ref{sect:networkml-intro} further discuss prominent tasks related to network analysis. Namely, Section \ref{sect:higher_order_network_analysis} discusses topics of capturing higher-order network structures called graphlets (subgraphs) in traditionally used pairwise graphs, which capture interactions between pairs of nodes, as well as shifting from pairwise graphs to hypergraphs, which are capable of capturing interactions between more than two nodes (InfoBox~\ref{box:terminology}). Section \ref{sect:networkml-intro} discusses machine learning advances in network biology, which has grown exponentially in the last decade. Key topics discussed include graph representation learning, incorporating knowledge into machine learning models, generative graph modeling, and transfer learning.

Section \ref{sect:personalized_medicine} complements the other, computationally focused sections by discussing an applied aspect of network biology: network-based personalized (or precision) medicine.  Precision medicine aims to provide tailored treatment strategies for individuals \cite{aronson2015building,kaiser2015nih}. This personalized characterization may include molecular, environmental, lifestyle, and other factors.  Integrating such different data types via network approaches can expand the potential for precision therapeutics while providing robustness to various types of data noise \cite{wang2014similarity}. 

The five topics are not mutually exclusive. For example, multimodal (including multi-omics) data integration is a topic relevant to almost all of Sections \ref{sect:network_inference_comparison}-\ref{sect:personalized_medicine}. 
After the current research network biology advances are presented in these five sections, Section \ref{sect:research_discussion_future_directions} discusses future research directions in the field, and Section \ref{sect:additional_discussion} provides additional discussion on scientific communities, education/training, and diversity in computational (including network) biology.

\begin{InfoBox*}[!t]
\fbox{
    \parbox{0.96\textwidth}{
    \begin{itemize}[]
        \item A (pairwise, homogeneous) \textbf{graph} (or network) $\mathcal{G} = (\mathcal{V}, \mathcal{E})$ is defined by a set of nodes (or vertices) $\mathcal{V}$ and a set of edges (or links) $\mathcal{E}$. All nodes $v \in \mathcal{V}$ are of the same type. 
        An edge $e_{u,v} \in \mathcal{E}$ indicates a relationship between exactly two nodes $u, v \in \mathcal{V}$. 
        \item In a \textbf{protein-protein interaction (PPI) network}, nodes are proteins and edges correspond to physical bindings between proteins. Such a network of physical PPIs is also referred to as \textbf{interactome}.
    
     \item {A (physical) PPI network is a special type of an \textbf{association network} between proteins. In addition to physical PPIs, an association network may contain links between proteins} derived from sequence or 3D structural similarities, genetic interactions, literature-mined edges, or other protein association types.  
    
     \item \textbf{Correlation networks} are calculated from -omics data collected across multiple samples. A prominent type are gene co-expression networks, where nodes (genes) are linked by undirected edges if the genes' expression levels are correlated strongly enough across the samples.
       
     \item \textbf{Regulatory networks} capture directed relationships between regulators and their targets and describe causal (rather than correlative) relationships between biomolecules. A prominent type are gene regulatory networks where the regulators are transcription factor proteins (or other molecules that impact gene expression, such as microRNAs) and the targets are genes.  
     
     \item \textbf{Biomedical knowledge graphs} describe semantic relationships between diverse biomedical entities (e.g., genes, diseases, and patients, as well as associated measurements). They represent facts using ``subject-predicate-object'' triples as the fundamental unit; the subject and object are nodes in the graph and the predicate (or relation) corresponds to a directed edge between the nodes.  

    \item A \textbf{condition-unspecific (or context-unaware) network} spans multiple conditions/contexts such as diseases, ages, cell types, tissues, etc., and ultimately, individuals. 
             
    \item A \textbf{condition-specific network} is inferred by integrating a context-unaware network with condition-specific node measurement (e.g., gene expression or mutation) data. The outcome of the data integration is identification of network regions that are ``active'' in the given condition, which can be seen as \textbf{condition-specific or disease-dysregulated pathways} (sparse, tree-like subnetworks) or \textbf{functional modules} (dense, clique-like subnetworks). 

    \item A \textbf{heterogeneous graph} contains multiple types of nodes and/or edges. 
        
    \item A \textbf{multiplex graph} is a heterogeneous graph with multiple edge types between the same nodes, possibly nodes of a single type, in which case the heterogeneity comes from the different edge types.   
        
    \item A \textbf{network-of-networks} is a heterogeneous graph in which different node types exist at different scales (or levels) and nodes at a higher level are graphs themselves at the lower level.
        
    \item \textbf{Multimodal data} that are represented as a  heterogeneous graph in network biology include \textbf{multi-omic} data such as epigenomic, transcriptomic, proteomic, and metabolomic molecular measurements as well as non-molecular data such as text and images from e.g., patients' electronic health records. 

    \item A \textbf{hypergraph}  is a generalization of a (pairwise) graph in which an edge (also called a hyperedge) can connect any number (including more than two) of the nodes.
        
    \item A \textbf{subgraph} (or subnetwork) $\mathcal{G}_S = (\mathcal{V}_S, \mathcal{E}_S)$ of a  graph $\mathcal{G} = (\mathcal{V}, \mathcal{E})$ consists of a set of nodes $\mathcal{V}_S \subseteq \mathcal{V}$ and a set of edges $\mathcal{E}_S \subseteq \mathcal{E}$ such that for each edge $e \in \mathcal{E}_S$, both of its end nodes must be in $\mathcal{V}_S$. 
        
    \item A subgraph is \textbf{induced} if and only if all edges between the nodes in $\mathcal{V_S}$ that exist in $\mathcal{E}$  are in $\mathcal{E_S}$. 
    
    \item \textbf{Graphlets} are connected, non-isomorphic, induced subgraphs of a (pairwise) graph. 
        
    \item \textbf{Hypergraphlets} are graphlet extensions from (pairwise) graphs to hypergraphs.
    
    \item A \textbf{cluster or community} in a graph is a set of topologically related nodes, typically densely connected to each other and loosely connected to nodes in other clusters.
    
    \end{itemize}
    }
}

\par
\caption{Basic terminology used in the paper. Note that distinct scientific communities in network biology, including graph theory, network science, data mining, machine learning, and artificial intelligence, may use varied terminology for the same concepts or identical terms for different concepts.\label{box:terminology}}
\label{infobox}
\end{InfoBox*}

%% file: 020.inference_comparison_rtf.tex
\begin{figure*}[t!]
  \centering
    \includegraphics[width=1\linewidth, trim=0cm 0.01cm 0cm 0cm]{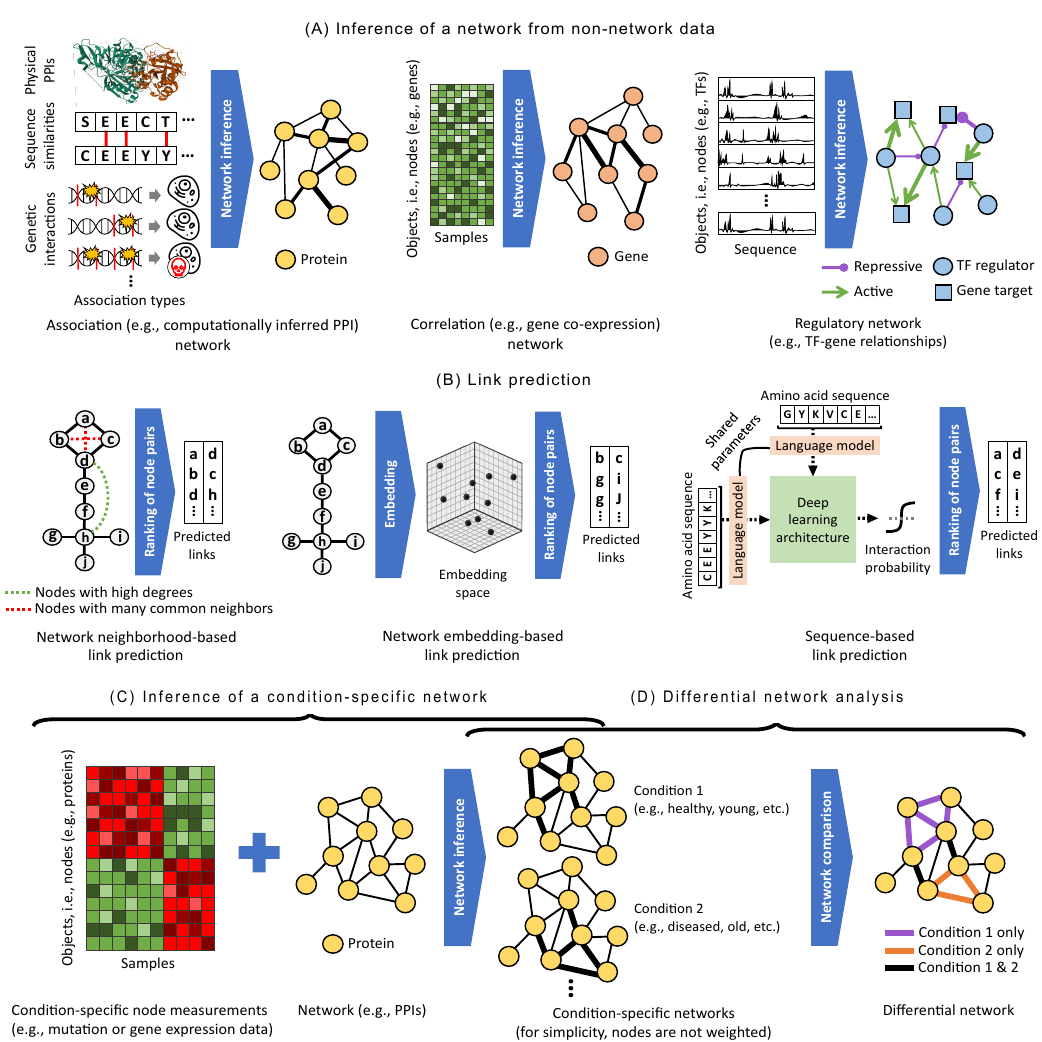}
    \caption{Prominent topics related to network inference and comparison. \textbf{(A)} Inference of an association (\emph{left}), correlation (\emph{middle}), or regulatory (\emph{right}) network from non-network data.  \textbf{(B)} Link prediction: inference of new interactions from existing network data via neighborhood- (\emph{left}) or embedding-based (\emph{middle}) approaches, or from sequence data (\emph{right}). For the former, shown are nodes that may be linked by new edges because two given nodes have high degrees (preferential attachment, green) or share many common neighbors (red); other neighborhood-based approaches exist, as discussed in the text.  \textbf{(C)} Inference of a condition-specific network. The second approach category is illustrated. The thicker an edge in the network for a given condition, the more relevant the edge is for that condition. \textbf{(D)} Differential network analysis. Illustrated is a potential differential network between conditions 1 and 2, in which the edges that are highly relevant for condition 1 but not condition 2 are in purple, and the edges that are highly relevant for condition 2 but not condition 1 are in orange; all other edges, which have consistent relevance patterns in both conditions, are shown in black.}
    \label{fig:inference_fig1}
\end{figure*}

\xhdr{Inference of a network from non-network data} 
Biological networks that are computationally inferred from non-network -omics data can be categorized into three broad types: association networks, correlation networks, and regulatory networks. {The three network types are defined briefly in InfoBox 1, discussed in detail in the following text, and illustrated in Fig. \ref{fig:inference_fig1}A.}
     
{Association networks typically capture  undirected and unsigned relationships between biological molecules; while they might contain experimentally derived interactions, they may also contain interactions derived computationally from a variety of possible data sources. One of the most common types of association networks are {(physical)} PPI networks, which are explicitly derived via high-throughput experiments (Section \ref{sect:introduction}) \cite{luck2020reference}.} {These experiments, primarily co-immunoprecipitation and yeast two-hybrid, differ in their estimated error rates and can produce both false positives and false negatives \cite{bader2004gaining,sprinzak2003reliable,von2002comparative}. In addition, for all but the yeast interactome, where a substantial fraction of pairs of proteins have been assayed, even in most model organisms, the majority of pairs of proteins have not been tested for interaction \cite{sledzieski2021d}. Thus, even across all the myriad sources of  PPI networks, there is much missing data~\cite{sledzieski2021d}.} {In addition to physical PPIs, many public resources curate associations between biomolecules from many data sources  \cite{bajpai2020systematic,wright2024state}}. For example, the {widely-used} STRING association network \cite{szklarczyk2023string} contains {interactions between proteins} derived from sequence or 3D structural similarities, genetic interactions, literature-mined edges, or other types of pairwise protein associations that are distinct from physical binding between proteins. 

{In an association network composed of genetic interactions (also known as} a genetic interaction network{)}, an edge between  nodes (genes/proteins) indicates that mutations or other perturbations to the two nodes produce an unexpected cellular phenotype \cite{baryshnikova2013genetic}. An example of a genetic interaction is when mutations in both of the genes/proteins result in cell death, i.e., are lethal, while the cell remains viable when there is a mutation in just one of them. A weighted version of a genetic interaction network also exists, in which edge weights indicate how strong or weak the observed double mutant phenotype, such as cell growth rate, is compared to the expected phenotype \cite{costanzo2016global}. 

{Challenges with association networks are that they are generally not condition-specific and also contain interactions derived for multiple different types of evidence, with different evidence sources having different quality levels and representing different types of biological relationships. Additional investigation of how different evidence sources influence network analysis results is often required \cite{Kim2021HumanNetv3}. Although the biological relationships represented in PPI networks and genetic interaction networks are easier to interpret, these networks tend to be incomplete and noisy and only exist for a limited number of species and biological conditions, limiting their use \cite{rolland2014proteome,Zitnik2019Evolution}.}
    
Correlation networks are typically calculated from -omics data collected across multiple samples (time points, tissues, patients, ages, drugs, or other conditions){; relationships in correlation networks are typically undirected and signed, depending on how the network is inferred}. Among the most prominent types of correlation networks are gene co-expression networks. Namely, given transcriptomics data containing the expression (i.e., mRNA abundance) levels of genes across multiple samples, a gene co-expression network can be constructed by linking nodes (genes) via  edges if the genes' expression levels are correlated strongly enough across the samples. In addition to being used to capture gene co-expression, correlation networks have been applied in biomedicine to study relationships between many other types of elements, such as metabolites \cite{perez2020network}, disease biomarkers \cite{chu2014analyzing,huang2019network,nishihara2017biomarker}, and even foods \cite{kim2015uncovering,samieri2020using}. Correlation networks are widely used in biomedical applications due to their simplicity and the ease with which they can be generated and interpreted \cite{huang2019network,lee2021changes,pierson2015sharing,samieri2020using}. Pearson correlation is the most common measure for calculating correlation networks, i.e., determining which gene pairs should be linked by edges, although other measures, such as Spearman correlation or mutual information, are also used, depending on the nature of the data and nonlinearity of the relationships being captured \cite{reshef2011detecting}. Multiple algorithms and tools have been developed for inferring correlation networks, including ARACNe \cite{margolin2006aracne}, which calculates the mutual information between pairs of nodes and then removes indirect relationships; CLR \cite{faith2007large}, which calculates the mutual information between pairs of nodes and then $z$-score normalizes; WGCNA \cite{zhang2005general}, which scales the Pearson correlation to generate a scale-free network topology (or network structure); and wTO \cite{gysi2018wto}, which normalizes the chosen correlation by all other correlations and calculates a probability for each edge.
    
{One advantage of correlation networks compared to association networks, especially PPI networks resulting from high-throughput experiments, is that correlation networks are explicitly derived from condition-specific -omics data, while association networks generally do not capture condition-specific information \cite{sonawane2019network}. However,} despite their popularity, correlation networks have multiple known limitations. One limitation is difficulty translating to biological mechanisms {\cite{larsen2019coli}}. Another limitation is that different network inference methods yield significant dissimilarities in the topology as well as functional content between the  resulting correlation networks \cite{rider2014}. For example, when multiple methods are applied to infer gene co-expression networks based on the same underlying data, the resulting networks tend to capture different sets of edges between the same nodes; furthermore, when those networks are used to predict genes' functional annotations such as Gene Ontology (GO) terms, the results often differ \cite{li2023enhancing}. Sometimes it might be helpful to combine networks inferred using different methods into a consensus network \cite{gysi2018wto,li2023enhancing}, where edges are re-weighted so that the more networks support an edge and the more strongly they support it, the higher its consensus weight or probability. A further limitation of gene co-expression networks is that co-expression between two genes occurs when one gene regulates another or when two genes are targeted by the same regulator {\cite{ku2012interpreting,yin2021emergence}}. However, these two distinct biological scenarios are represented the same way in a co-expression network, by linking the two genes with an undirected edge. Instead, regulatory networks can distinguish between the different scenarios, as discussed next.

Regulatory networks capture directed relationships between regulators and their targets and describe causal (rather than {just} correlative) relationships between biomolecules{; although these networks in theory should be signed, in practice deriving the sign of regulatory relationships from high-throughput biological data is challenging}. There are many types of regulatory networks in biology. However, for most inferred regulatory networks, the regulators are transcription factor (TF) proteins (or other molecules that impact gene expression such as microRNAs) and the targets are genes; these are commonly referred to as gene regulatory networks. There are many approaches to infer gene regulatory networks. For example, TF-gene relationships can be measured experimentally through ChIP-sequencing. In this case, the presence of a TF binding in the regulatory region(s) of a gene can be used to infer an edge from that TF to the gene. However, the cost and experimental limitations make it impossible to infer a complete gene regulatory network in this way. Therefore, many computational approaches have been developed to infer gene regulatory networks. For example, the DNA sequence of gene regulatory regions can be scanned to identify matching patterns (known as sequence motifs) that indicate a potential TF binding site; however, linking TFs to genes based on DNA sequence alone does not give a condition-specific network. Thus, methods to infer gene regulatory networks typically use gene expression data, either alone or in combination with computational evidence for TF binding in gene promoters, to infer TF-gene relationships \cite{marbach2012wisdom}. Popular algorithms of this type include Inferelater \cite{bonneau2006inferelator}, which uses linear regression, L1 shrinkage, and LASSO to identify a set of parsimonious models to predict target gene expression levels from TF expression levels (and other factors); GENIE3 \cite{huynh2010inferring}, which uses tree-based ensemble methods to develop a set of regression problems that predict the expression pattern of each target gene from the expression of a set of input TF genes; and PANDA \cite{glass2013passing}, which uses message passing to amplify consistent structures across three input data types: TF-TF PPIs, computationally inferred TF-gene relationships, and gene-gene co-expressions. As opposed to Inferelator and GENIE3, PANDA does not consider the expression levels of TFs but instead uses evidence of co-expression in genes as evidence of targeting by the same TF. In contrast, a recent method NETREX-CF incorporates, among other techniques, a machine learning approach known as collaborative filtering to deal with missing data \cite{wang2022netrex}.
    
Other methods to infer regulatory networks incorporate epigenetic data. In particular, chromatin state can indicate whether the DNA is ``open'' and available to be bound by a TF; thus, computational evidence for TF binding in gene regulatory regions that also overlap with open chromatin can be used to estimate cell type-specific networks \cite{neph2012circuitry}. Specific algorithms to infer gene regulatory networks using epigenetic data include TEPIC \cite{schmidt2017combining,schmidt2019tepic}, which combines TF binding affinities, chromatin state data, and gene annotation data to predict TF-gene relationships, and SPIDER \cite{sonawane2021constructing}, which uses message passing to infer and amplify consistent structure in an epigenetically-pruned gene regulatory network constructed by combining computational evidence for TF binding with open chromatin data. Both TEPIC and SPIDER can also (optionally) incorporate gene expression data. Despite multiple methods in this area (including many beyond those described here), it remains challenging to integrate multiple types of -omics data to effectively infer accurate condition-specific regulatory networks; {we elaborate on this challenge in subsection ``Inference of a heterogeneous network from multimodal data'' of Section \ref{sect:multimodal_networks_integration}}.

\xhdr{Link prediction: inference of new interactions from existing network data}
{Link prediction is applicable to any network type, but in network biology, it has prominently been used in association networks containing interactions between proteins. Regardless of the type of data used to construct an  association network, the resulting network is often incomplete. For example, many pairs of proteins in an organism may yet to be assayed for physical interaction. However, the ``guilt by association'' principles that underly the topological organization of most of these networks~\cite{cowen2017network} mean that the patterns of connection of existing links can reliably predict some of the missing edges. We refer to this as network-based link prediction. }
Network-based prediction of new {interactions between proteins} often uses either a relatively simple rule (e.g., it may be desirable to link nodes that have high degrees, that have many common interacting partners -- or neighbors -- either direct or extended ones, that share many paths, or that are topologically similar \cite{Hulovatyy2014}) or more sophisticated diffusion-based network embeddings \cite{cocskun2021node,cowen2017network,devkota2020glide,hamilton2018embedding,huang2020skipgnn,kovacs2019network,Yuen2020better}. A mixture of these strategies, where simple rules are employed in the core of the network, and diffusion-based network embeddings are employed outside the core, perform particularly well. However, the set of rules and the embedding used matters \cite{devkota2020glide}, especially because interaction patterns may be quite different in {networks containing physical PPIs vs. those containing inferred, non-physical associations between proteins}.  

\xhdr{{Link prediction: other techniques to infer missing interactions}}
{Beyond methods that leverage the topology of the known interactions, the other methods to infer missing interactions will vary based on the underlying type of {protein} association data used to construct the network. For example, for physical PPI  prediction, classical techniques such as docking can also be used when protein 3D structural models are available. With the rise of deep learning methods such as AlphaFold~\cite{jumper2021highly}, ESMFold \cite{lin2022language}, and OmegaFold \cite{wu2022high}, now a 3D structural model is usually available for most proteins. AlphaFold-Multimer~\cite{evans2021protein} is a recent deep learning-based extension of AlphaFold that allows for predicting protein complexes, i.e., the quaternary structure of multiple proteins; then, it might be possible to use the confidence score of the predicted structure to predict whether the proteins interact or not. The predicted quaternary structure also provides the interaction interfaces between the proteins.}

{When the goal is ultra-fast prediction (for example, in order to perform genome-wide scans), there are alternative  deep learning methods} \cite{chen2019multifaceted,hashemifar2018predicting,sledzieski2021d,zhang2018predicting} {that} have had success in sequence-based prediction of PPIs. These methods focus on computational speed. That is, like the network-based methods, they seek to predict {only} \emph{whether} (rather than also \emph{how}, which is more challenging) two protein sequences interact, so that it is tractable to make predictions for all the protein pairs in the network.  However, {we note that some of these} sequence-based methods manage to implicitly incorporate information about protein 3D structures. For example, D-SCRIPT \cite{sledzieski2021d} uses a pretrained protein language model  \cite{bepler2021learning} and implicitly learns a fuzzy contact map representation.

How to simultaneously leverage network- and sequence-based link prediction {for physical PPI data} remains an open problem,  with valuable initial work \cite{bepler2021learning}. Also, evaluating link prediction methods and especially hybrid methods is tricky. This is because existing ground-truth  networks (other than HURI \cite{luck2020reference}) are biased by the portions of the networks containing well-studied proteins and pathways {\cite{Schaefer2015correcting}}. So, it is difficult to come up with fair performance measures that are not biased by node degrees, and that do not advantage network-based methods while disadvantaging sequence-based methods {\cite{Singh2022TopsyTurvy,Wang2023assesment}}. On the other hand, sequence-based approaches do better on close homologs of known interacting protein pairs {\cite{Sledzieski2023TT3D}}.

{Other researchers have noted that databases that amalgamate physical PPI data have not always kept up with the literature, and have proposed text-mining approaches to predict these ``missing'' links \cite{kim2008pie,papanikolaou2015protein,van2009novel}.}

\xhdr{Inference of a condition-specific network} 
While existing biological network data resulting from extensive experimental efforts are an incredible resource, they {typically} do not capture how interactions in biological networks differ across conditions, i.e., they are context-unaware. By conditions, we mean diseases, ages, cell types, tissues, etc., and ultimately, individuals. Indeed, while human genomes in both healthy and disease populations are rapidly being sequenced, the corresponding condition-specific networks remain largely unknown. Moreover, the substantial amount of genetic variation across populations makes it infeasible in the near term to experimentally determine the full impact of this variation on interactions. So, computational methods have played and will continue to play a major role in inferring condition-specific networks.

We divide computational approaches for inferring condition-specific networks into several broad categories{. (1) The first category are} approaches that assess whether mutations observed in disease alter protein interactions{. (2) The second category are} approaches that combine mutation data (e.g., on how many patients with a disease have genes containing significantly associated single nucleotide polymorphisms, indels, etc.) or condition-specific gene expression data  (e.g., information on which genes are significantly expressed -- or active -- in a given condition (here, typically multiple data samples are needed per condition) with a PPI network. {The goal is}  to identify PPIs that are dysregulated in a given disease or active in a given condition, i.e., to infer a condition-specific PPI network  (Fig. \ref{fig:inference_fig1}C){. (3) The third category are} approaches that use gene expression data to infer a correlation network specific to the condition or sample of interest. {(4) The fourth category are} analogs of the previous approaches but applied to regulatory networks rather than PPI or correlation networks.

Regarding the first approach category, significant computational efforts have focused on characterizing whether mutations observed in disease and variants across populations alter protein interactions. Early work mapping mutations observed in Mendelian diseases onto protein structures demonstrated that there is a statistically significant enrichment of Mendelian disease mutations in protein interaction interfaces, as compared to neutral polymorphisms observed across populations \cite{gao2015insights}. Homology modeling and domain-based approaches to identify sites that participate in interactions with DNA, RNA, peptides, ions, and small molecules have revealed that missense mutations observed in Mendelian diseases and somatic missense mutations in cancer are both enriched in these sites, with the strongest enrichments for DNA-binding sites, while common variants are depleted from these sites \cite{ghersi2014interaction,kobren2019systematic}. Further, these enrichments can be leveraged to identify cancer-relevant genes by developing statistical approaches to uncover proteins with more somatic missense mutations in their binding sites than expected \cite{ghersi2014interaction,kobren2020pertinint}. Protein interaction interfaces, as identified by homology modeling \cite{mosca2013interactome3d} and machine learning \cite{meyer2018interactome}, have also been shown to be enriched in somatic missense mutations as compared to non-interface residues, and specific protein interactions relevant for cancer have been identified \cite{cheng2021comprehensive}. High-throughput experimental screens have led to estimates that two thirds of disease-causing polymorphisms perturb protein interactions, with about half of these interrupting specific protein interactions while leaving other interactions unaffected \cite{sahni2015widespread}.

Regarding the second approach category, numerous computational efforts have focused on integrating condition-specific molecular measurements, mainly gene mutation or expression data (also referred to as gene activity data), with PPI network data (which is generally not condition-specific, i.e., is context-{unaware}){. They do so} by mapping the gene activities onto the corresponding proteins in the PPI network, in order to assign condition-specific weights to the proteins or PPIs (or both) in the network (Fig. \ref{fig:inference_fig1}C){. Then,} highly weighted PPI network regions are hypothesized to be pathways dysregulated in disease (if using mutation data) or condition-specific subnetworks (if using expression data) \cite{Leiserson2015,newaz2020improving}. The set of all such PPIs/pathways/subnetworks is a condition-specific PPI network.  The data integration step is often performed via network propagation \cite{cowen2017network}, which diffuses the gene activities through the PPI network via random walks{. Nonetheless,} other approach types exist such as kernel, Bayesian, or non-negative matrix factorization methods \cite{newaz2020improving}.

{Prominent applications of approaches from the second category have been studying cancer \cite{Leiserson2015,silverbush2019simultaneous}}, tissue-specificity \cite{Basha2020}, aging \cite{li2022towards}, and genome-wide associations \cite{carlin2019fast,vanunu2010associating}. As an example, cancer-related gene mutation data was integrated with PPI data using the HotNet2 algorithm to identify the parts of the PPI network that are likely to be active in cancer \cite{Leiserson2015}. Such a cancer-specific network is not  necessarily connected, i.e., it might consist of multiple connected components, each of which can be thought of as a cancer-specific pathway or subnetwork. As another example, a general framework was proposed for assessing the ability of condition-specific PPI network inference approaches to illuminate tissue-specific processes and disease genes \cite{Basha2020}. This framework integrated RNA-sequencing profiles for 34 human tissues with a PPI network to create 34 tissue-specific PPI networks. Here, all tissue-specific PPI networks contained the same nodes and interactions, and they differed ``only'' in the weights associated with them. Then, given data associating GO  biological processes to their relevant human tissues, this framework allows different condition-specific PPI network inference approaches to be benchmarked via enrichment tests in terms of their ability to recover tissue-specific processes. As a final example, unlike in the above applications where the inferred cancer- and tissue-specific networks were static, when studying human aging, which is a dynamic biological process, it is desired to infer a dynamic aging-specific network. Of the pioneering approaches towards this goal \cite{li2020supervisedTCBB,newaz2020improving,li2021improved,li2022towards}, a recent finding is that inferring an aging-specific PPI network that is both weighted and dynamic (as opposed to unweighted or static) results in the most accurate prediction of aging-related genes \cite{li2021improved}. To infer this network, network propagation was used to map gene expression-based weights at different ages onto nodes in a PPI network. This resulted in a weighted network snapshot for each age, where the different snapshots had the same nodes and PPIs and ``only'' differed in their age-specific weights. The collection of all age-specific snapshots formed a weighted dynamic aging-specific PPI network. Then, aging-related genes can be predicted from this network, as discussed below \cite{li2021improved,li2022towards}.

An important issue in identifying condition-specific networks and especially disease-altered subnetworks via the above approaches is to determine whether the resulting (sub)networks are due to the molecular measurements (i.e., mutation or expression data) alone, the PPI network topology alone (e.g., due to ascertainment bias in PPI network data), or genuinely a combination of both molecular measurement and network data. Recent work has shown that in some applications there may be a narrow regime where both molecular data and network information contribute to the identification of disease-dysregulated subnetworks \cite{chitra2022netmix2,reyna2021netmix}.

Regarding the third approach category, condition-specific correlation networks are most often derived by applying a correlation measure to subsets of related samples \cite{pierson2015sharing}. However, since correlation measures rely on defining a distribution, this approach is inappropriate when a specific condition is represented by only a few (or even a single) sample(s). However, recently methods have been developed to infer ``sample-specific correlations''. That is, given a set of gene expression samples (across which correlation can be measured), these approaches can estimate one network for each individual sample in the input dataset. In particular, both SSN \cite{liu2016personalized} and LIONESS \cite{kuijjer2019lionessr,kuijjer10estimating} work by computing two correlation networks, one with all samples and one with all samples except an individual sample of interest{. Then, they}  use the difference between the two {networks} to estimate a correlation network specific to the sample of interest.

Finally, regarding the fourth approach category, genetic variants can impact gene regulatory networks by, for example, altering TF binding or allele-specific expression \cite{przytycki2020differential}. Recall that missense mutations are enriched in sites that participate in interactions with DNA, RNA, peptides, ions, and small molecules, with the strongest enrichments for DNA-binding sites \cite{ghersi2014interaction,kobren2019systematic}{. Also, recall} that statistical approaches to identify proteins with more somatic missense mutations in their binding sites than expected by chance have identified cancer-relevant genes \cite{kobren2020pertinint,kobren2019systematic}. Deep learning approaches trained on DNA binding data from ENCODE \cite{encode2020expanded} have also been used to assess whether DNA mutations impact TF binding in a tissue-specific manner \cite{Zhou2015}. For some TFs, altered DNA-binding specificities can be predicted \emph{de novo} using machine learning \cite{christensen2012recognition,persikov2014novo,sahni2015widespread,wetzel2022learning}. However, if a DNA-binding protein's specificity is known \emph{a priori}, then it is more accurate to instead predict how mutations alter that specificity rather than predict specificities \emph{de novo}. For example, accurate predictions about how mutations alter DNA-binding specificities for homeodomain proteins were made by simultaneously learning interaction interfaces between DNA-binding proteins and their binding sites together with a predictive approach for DNA-binding specificity \cite{wetzel2022learning}. Extending this approach to all DNA-binding proteins represents an important avenue for future work.

There has been also significant work done to infer condition-specific regulatory networks from various types of -omics data, as has been extensively reviewed in \cite{baur2020data}. As one example, PANDA was applied to subsets of GTEx gene expression data to infer 38 tissue-specific gene regulatory networks \cite{sonawane2017understanding}{; then, it} was found that changes in TF targeting patterns led to the creation of new regulatory paths, giving them transcriptional control of tissue-specific processes. There  also exist approaches that can be used to infer sample-specific networks for different -omics data types. For example, EGRET integrates predicted TF binding sites with genotype and expression quantitative trait loci data to create individual genotype-specific regulatory networks \cite{weighill2022predicting}. The SPIDER \cite{sonawane2021constructing} and TEPIC \cite{schmidt2017combining,schmidt2019tepic} methods (described above) can be applied to individual epigenetic profiles to generate sample-specific regulatory networks. PSIONIC learns patient-specific TF regression weights by using chromatin-filtered TF-gene relationships to predict gene expression. Finally, the LIONESS method \cite{kuijjer10estimating} can be used together with existing gene regulatory network reconstruction approaches that leverage gene expression data. When applying it in the same way as already described for correlation networks (the third approach category above), the LIONESS framework uses two estimated gene regulatory networks, one inferred with all gene expression samples and one inferred with all samples except one, to estimate a gene regulatory network specific to that sample \cite{kuijjer10estimating}.

\xhdr{Differential network analysis: comparison of  condition-specific networks} 
Condition-specific networks often have the same set of nodes and differ {only} in terms of their edges. Many approaches have been developed to identify network regions that differ the most between condition-specific networks; {such regions have been shown to be} responsible for the underlying biological differences between  e.g., healthy and disease conditions, between different tissues, or between young and old ages  \cite{basha2018differentialnet,lichtblau2017comparative}{, as discussed  in more detail below}. In general, approaches for this task can be characterized in several ways.

One category is based on the stage of network analysis, i.e., \emph{when} differences between condition-specific networks are measured. Given condition-specific networks, one option is to first compute some topological property of a network region (at the level of a node, edge, network cluster --  group of highly interconnected nodes -- or entire network; see below) in each condition-specific network and then measure the extent of change in that property across the networks/conditions; the goal is to identify network regions that change the most \cite{lichtblau2017comparative,zhu2016metadcn}. {By a topological property, we mean a quantifiable measure of network structure such as the degree distribution of a network (the percentage of nodes in the network that have a given number of neighbors, i.e., degree), or centrality measures that rank nodes in a network from most to least central/important (examples are degree centrality according to which nodes with high degrees are central, and betweenness centrality according to which nodes that are on many shortest paths are central) \cite{Barabasi2016,newaz2019bookchapter,newman2018networks}.}

{A potential issue is that some topological properties, and especially centrality measures,} are meaningful when used within a network but not necessarily when compared across networks \cite{newman2018networks}. As an alternative, approaches exist that first use the condition-specific networks to infer a single differential network that intuitively captures edges that differ between the conditions  (Fig. \ref{fig:inference_fig1}D); only then, a desired topological property (e.g., centrality of each node) in the differential network is computed to identify network regions that are the most relevant (e.g., central/important) for the underlying condition-specific differences \cite{ruan2015differential}.

The other category is based on the level of topology, i.e., \emph{where} differences between condition-specific networks are measured: at the node \cite{weighill2021gene}, edge \cite{glass2015network}, cluster \cite{padi2018detecting}, or entire network level \cite{newaz2020improving}. At the node level, differences in centrality (e.g., degree or betweenness) are often used to identify the biomolecules around which network connectivity varies the most between the compared conditions. For example, ``differential targeting,'' i.e., the difference in gene targeting -- or the sum of the weights for all incoming edges to a gene --  between two gene regulatory networks was used in combination with standard gene set enrichment tools to identify over-represented biological processes in pancreatic ductal adenocarcinoma subtypes \cite{weighill2021gene}. At the edge level, the goal is typically to determine edges specific to a given condition. This can be done in multiple ways, by taking, for example, a certain percentage of the highest-weight edges, all edges above a given threshold, edges that have higher weights in one condition compared to others \cite{sonawane2017understanding}, or a combination of these \cite{glass2015network}. For example, the tissue-specific PPI networks discussed above, which were defined by differential edge scores, were correctly enriched in their respective tissue-associated biological processes; also, when the top 1\% of the differential edges were considered, the resulting differential network regions were correctly enriched in genes related to diseases associated with their respective tissues \cite{Basha2020}. Linking this discussion to the first approach category described above, it is important to note that although node centralities are often determined for each condition-specific network and then compared across the networks, they  can also be calculated for a network defined by condition-specific edges. For example, degree and betweenness centralities of all genes in 38 tissue-specific gene regulatory networks were used to show that tissue-specific genes tended to assume bottleneck positions in their corresponding networks; in parallel, tissue-specific edges were identified by comparing the weight of each edge in a given tissue to the distribution of that edge's weight across all tissues, and it was found that the tissue-specific edges were enriched for connections between tissue-specific genes and depleted for canonical interactions  \cite{sonawane2017understanding}. At the cluster level, for example, given two condition-specific networks, ALPACA \cite{padi2018detecting} identifies clusters that are shared between networks and distinct to each network. Heterogeneous (specifically, multiplex; Section \ref{sect:multimodal_networks_integration}) clustering algorithms \cite{mucha2010community} could also perhaps be useful for identifying such clusters. At the level of entire networks, typically their pairwise edge overlaps, as measured by e.g., the Jaccard index, are used to quantify their pairwise (dis)similarities \cite{newaz2020improving}.

We comment on two additional aspects of differential network analysis. First, while some condition-specific networks are derived from multiple data samples, sample-specific networks have the additional benefit of being able to be compared while accounting for other potentially relevant biomedical information \cite{kuijjer10estimating}. For example, the same statistical tools employed for differential gene expression analysis can be used to determine significant changes in the node-, edge-, cluster-, and network-level topological properties between sets of sample-specific networks. Importantly, this allows topological properties to be evaluated in the context of relevant biological and phenotypic variables, as well as potential confounders. For example, limma \cite{ritchie2015limma} was applied to compare features between male and female sample-specific gene regulatory networks while controlling for relevant confounders such as body mass index and age; node, edge, and TF-targeting was identified specific to males and females across 29 different tissues \cite{lopes2020sex}, as well as sex-specific targeting of the drug metabolism pathway in colon cancer \cite{lopes2018gene}. 

Second, while the above discussion applies to all condition types, including temporal ones, we explicitly wish to comment more on approaches for characterizing how networks change over time \cite{teschendorff2021statistical}. A prominent application in this context has been studying the change of PPI network topology with age. The process of inferring an aging-specific PPI network has already been discussed above. Here, we comment on how such a network, consisting of network snapshots corresponding to different ages, is analyzed. Original studies asked whether the overall, or global,  topology changed with age, by: measuring pairwise edge overlaps between the snapshots; evaluating whether the snapshots' properties such as the average clustering coefficient, diameter, and graphlet degree distributions changed with age; and evaluating the fit of each snapshot to random (e.g., scale-free or geometric) graphs \cite{faisal2014dynamic,newaz2020improving}. Global topologies of the age-specific snapshots did not significantly change with age. It was then analyzed whether local topological positions of nodes as measured by (normalized) centralities changed with age. Hundreds of such genes were identified and predicted as aging-related; the predictions were validated via functional enrichment analyses  \cite{faisal2014dynamic,newaz2020improving}. 

{Unlike} such unsupervised prediction of aging-related genes, in recent work \cite{li2021improved,li2022towards},  supervised prediction was performed: by relying on knowledge about which genes are aging- versus non-aging-related \cite{Magalhaes2009a}, new aging-related genes were predicted if their evolving topologies in a dynamic aging-specific PPI network matched topologies of the known aging-related genes. Recall that the state-of-the-art aging-specific dynamic PPI network is weighted. So, weighted node topological measures were used as features for supervised prediction that were simple extensions of unweighted centralities. Also, more advanced measures were proposed, which account for how the distribution of edge weights in the given node's (extended) network neighborhood changes with age, i.e., across the network snapshots \cite{li2021improved}. A parallel line of work focused on studying how clusters, i.e., community structure,  in a dynamic aging-specific human PPI network changed with age, and it was shown that the most prominent changes in the community structure correspond to ages that reflect known shifts from one stage of human lifespan to another \cite{ClueNet,SCOUT}. 

Another prominent point of discussion in the temporal/dynamic context are theoretical studies of molecular networks and observations of cell differentiation (i.e., the transition of a cell from one type to another), which indicate that cellular transitions can be smooth or nonlinear, gradual or abrupt \cite{moris2016transition,nykter2008gene}. Computational methods to characterize these transitions using single-cell gene expression data include MuTrans \cite{zhou2021dissecting}, QuanTC \cite{sha2020inference}, and BioTIP \cite{yang2022detecting}. These methods use different statistical approaches (stochastic differential equations, unsupervised learning of cell plasticity, or co-expression) and underlying theories (entropy and energy or tipping-point theory), but converge at the same best-studied bifurcations in six datasets \cite{yang2022detecting}. 

\xhdr{Other types of network comparison} 
Differential network analysis is one type of network comparison, in which networks being compared have the exact same nodes and differ ``only'' in their edges (or edge weights). In other words, the mapping between the nodes of the compared networks is known. A complementary category of network comparison includes approaches that compare networks when their node mapping is unknown. Here, there are two distinct types: (1) network alignment or alignment-based network comparison and (2) alignment-free network comparison \cite{Yaverouglu2015}. 
 
Alignment-based network comparison aims to find a mapping between the nodes of the compared networks that optimizes some objective function; this typically means conserving many edges and a large subgraph between the networks \cite{faisal2015post,Guzzi2017,Yaverouglu2015}. This approach category is useful for comparing biological networks of different species to identify evolutionary conserved parts of the networks. Consequently, network alignment allows for transferring biological knowledge (e.g., proteins' functional annotations or PPIs)  between aligned network regions across the compared species; also, it can complement sequence alignment by allowing for identification of protein orthology relationships based on the proteins' PPI network  rather than (just) sequence similarities. Note that even when aligning homogeneous networks, the problem of network alignment can be viewed as integrating these networks into a heterogeneous (specifically, multiplex; Section \ref{sect:multimodal_networks_integration}) network representation. For this reason, and because recently methods have been proposed that actually align heterogeneous networks, we discuss algorithmic aspects of network alignment in the more appropriate Section \ref{sect:multimodal_networks_integration}. Here, we {mainly} aim to contrast general working principles of the different types of network comparison.
 
In contrast to alignment-based comparison, alignment-free network comparison {simply} aims to quantify the overall topological similarity between networks, regardless of a node mapping between the networks, and without intending to identify any conserved network regions; this typically means comparing some topological properties between networks, such as their (graphlet) degree distributions \cite{newaz2019bookchapter,Yaverouglu2015}. Alignment-free network comparison is most often used to evaluate the fit of a random graph (e.g., scale-free or geometric) to a real-world network; also, it can identify groups/families of networks that are topologically similar to each other \cite{Yaverouglu2015}. Given that alignment-free network comparison approaches do not aim to produce a node mapping between the compared networks, while alignment-based approaches do, the former are typically computationally more efficient than the latter \cite{Yaverouglu2015}.

%% file: 030.multimodal_rtf.tex
\begin{figure*}[t!]
  \centering
    \includegraphics[width=1\linewidth, trim=0cm 0.3cm 0cm 0cm]{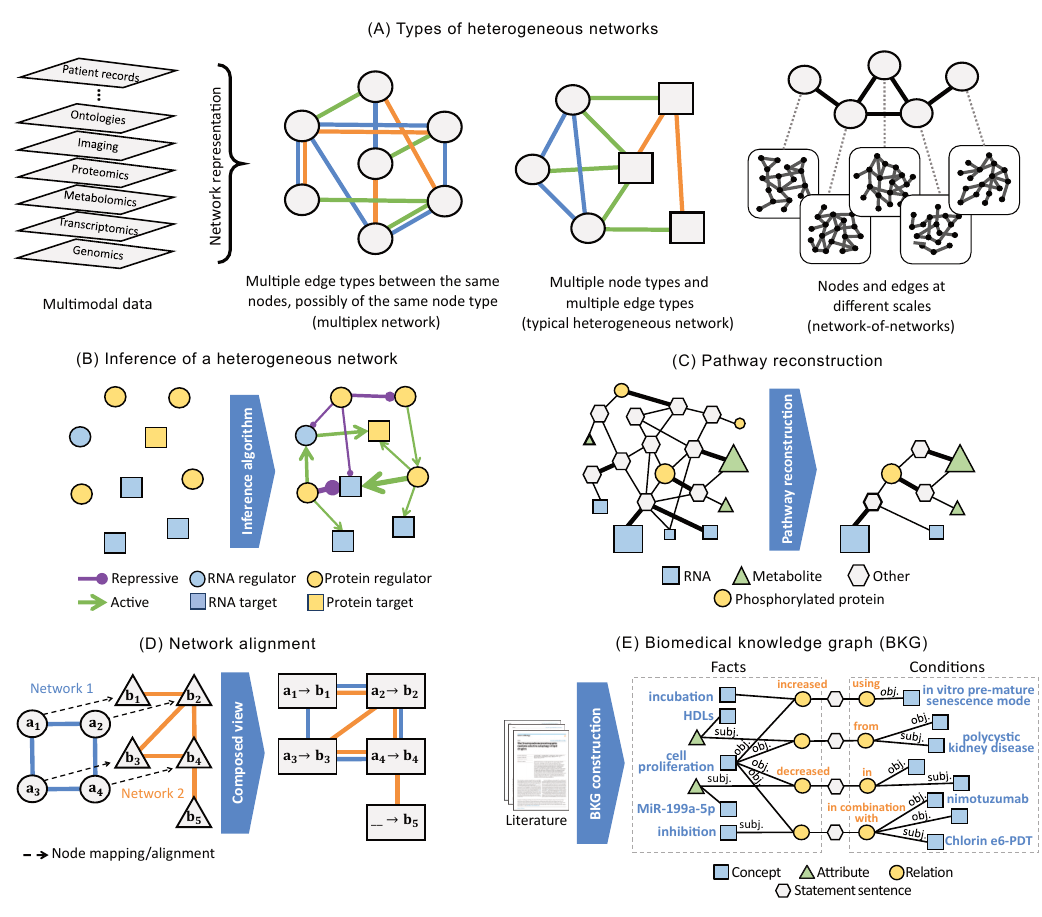}
     \caption{Prominent topics related to multimodal data integration and heterogeneous networks. \textbf{(A)} Heterogeneous networks can naturally represent multimodal data. A heterogeneous network can have only a single node type, with different data modalities representing multiple edge types. Or, there can exist both multiple node and edge types. Different node types can exist at different biological scales; e.g., in a network-of-networks, nodes at a given scale are networks at the lower scale.  \textbf{(B)-(E)} Prominent topics related to heterogeneous networks. \textbf{(B)} Inference of a heterogeneous network aims to learn the graph topology from multimodal -- to date, typically multi-omic --  measurements. \textbf{(C)} Pathway reconstruction for interpretation of multi-omic data: the input is multi-omic data and a background molecular network, and the output is a sparse subnetwork. Typically input biomolecules with higher scores (indicated by node sizes) and higher-quality connections (indicated by edge thickness) are prioritized in the output. \textbf{(D)} Network alignment: input can be individual homogeneous networks (\emph{left}) or heterogeneous networks. Even alignment of homogeneous networks leads to a heterogeneous network (\emph{right}) whose ``supernodes'' contain mapped nodes and whose edge types indicate which edges of the original networks are conserved (blue and orange between two ``supernodes'') versus non-conserved (blue only or orange only between two ``supernodes'') under the given node mapping. \textbf{(E)} Inference of and reasoning on BKGs. Shown is a condition-aware BKG. The middle nodes (hexagons) are statement sentences. The layers on their left represent fact tuples and those on their right represent the conditions associated with the facts. The tuples have relation nodes (circles), concept nodes (squares), and optional attribute nodes (triangles).}
     \label{fig:multimodal_fig1}
\end{figure*}

\xhdr{Overview} 
Network representations of biological systems, from cells to ecosystems, are naturally heterogeneous, consisting of multiple types of nodes and interactions \cite{de2023more}. This section focuses on prominent computational challenges related to inference and analysis of heterogeneous networks. Broadly, a heterogeneous network is defined as a representation of multimodal data where each data mode corresponds to a different node or edge type. In the literature, the term ``heterogeneous network'' has often been used as a synonym to, e.g., a multiplex, interdependent, multiscale, or multilayer network. The challenge is that sometimes different terminologies are used for the same concept, or the same terminology is used for different concepts; the disparate terminology associated with heterogeneous networks can reflect nuances in their frameworks \cite{kivela2014multilayer}. Here is the terminology from the existing literature (e.g., \cite{gu2022modeling,pio2021multiverse}) that we use in this paper (Fig.~\ref{fig:multimodal_fig1}A). 

A heterogeneous network is a network with multiple node types and/or multiple edge types. A multiplex network is a special type of heterogeneous network with multiple edge types between the same nodes, possibly nodes of a single type, in which case the heterogeneity comes from the different edge types. A multiplex network can be viewed as being composed of different network layers sharing the same set (replica) of nodes but each layer having distinct  edge types \cite{kinsley2020multilayer}. An example of this type in biology is a molecular network capturing different types of relationships, such as physical interactions, functional relationships, and sequence similarities between proteins. A typical heterogeneous network, including those discussed in this section, contains both distinct node types and (by definition) distinct edge types. An example of this type is a molecular network representing relationships among heterogeneous node types such as genes, transcripts, proteins, and metabolites. Another example is a knowledge graph representing semantic relationships between node types such as genes, patients, drugs, and diseases. Another level of complexity is handling distinct node types at different scales (or levels) of biological organization, e.g., node types resulting from data modalities that capture molecular measurements in epigenomic, transcriptomic, proteomic, and metabolomic assays and from non-molecular text and imaging data. Here, a network-of-networks is a special case in which a node at a given scale is a network at the lower scale. For example, a node (protein) in a PPI network can be represented as a protein structure network in which nodes are the protein's amino acids and edges link amino acids that are close enough in the protein's 3D fold \cite{gu2022modeling}. 

The broad definition of a heterogeneous network that we use subsumes any network type that is not a homogeneous (single node type and single edge type) network. Note that in some scientific fields, such as physics, while a multiplex network typically has the same meaning as above, heterogeneous network is a rarely used term. Instead, a heterogeneous network is often referred to as a multilayer network, and a network-of-networks is sometimes used as a synonym for a multilayer network \cite{de2023more,DeDomenico2013Mathematical,kivela2014multilayer}.

Heterogeneous networks are a powerful framework for the representation, integration, and analysis of diverse data modalities of a complex system with multiple types of nodes or edges (or both), allowing for reconciling complementary measurements and providing a holistic view of the system. Here, we discuss the following major research directions encompassing heterogeneous networks: inference of a heterogeneous network from multimodal data, pathway reconstruction for interpretation of multi-omic data, network alignment, inference and reasoning with biomedical knowledge graphs, and network-of-networks analysis. This is not an exhaustive list of topics on heterogeneous networks, and other sections touch on additional topics. For example, Section \ref{sect:networkml-intro} touches on graph representation learning including but not limited to learning in heterogeneous networks, and Section \ref{sect:personalized_medicine} talks about integration of multimodal data for the purpose of patient stratification, identification of disease-dysregulated molecular pathways and functional modules, and other precision medicine applications.

\xhdr{Inference of a heterogeneous network from multimodal data}
Heterogeneous network inference is the computational task of inferring the graph connectivity structure from multimodal -- to date, typically multi-omic --  measurements \cite{hawe2019inferring}.  The vast majority of methods for this task infer connections between nodes corresponding to biomolecules such as genes, proteins, and metabolites (Fig. \ref{fig:multimodal_fig1}B) using bulk -omic datasets. Single-cell -omic datasets have posed new opportunities for network inference where nodes can represent individual cells. Heterogeneous network inference methods can be grouped into categories based on how much they rely on labeled positive examples of edges. 

Probably the simplest category of approaches take as input labeled examples of edges and non-edges along with pairwise node feature vectors derived from multimodal data and train binary classifiers to discriminate node pairs with edges from node pairs without edges \cite{greene2015understanding,marbach2012predictive}. These binary classification approaches assume that all node pairs are independent of each other and are therefore limited in their ability to exploit the known connectivity structure of the graph. An alternative are embedding methods (discussed in more detail in Section \ref{sect:networkml-intro}) that take as input an incomplete graph and multimodal measurement data as node features and learn an embedding of the nodes based on the (partial) graph structure and measured values, which are then used to infer edges based on link prediction \cite{lee2020heterogeneous,yue2020graph} or matrix completion \cite{natarajan_inductive_2014}. Graph embedding methods relax the independence assumption of binary classification methods. As graph embedding methods capture more of the network connectivity, it is conceivable that they need less training data to do as good prediction as simple binary classification. Graph neural networks (GNNs, discussed in more detail in Section \ref{sect:networkml-intro}) offer new ways to incorporate more global information about the network to inform the inference task \cite{yue2020graph}. The biggest limitation of the above approaches is the need for positive training data (edges) and that negative examples (non-edges) are not truly observed but are assumed to be part of the complement of the positive set. 	

On the other hand, unsupervised graph structure learning methods take as input node-level measurements and infer the graph structure from these measurements alone, without requiring any labeled examples of edges/non-edges. These approaches can range from correlation-based networks inferring pairwise dependencies between nodes representing different multimodal data \cite{vasaikar2018linkedomics,zhou2021omicsanalyst} to more general approaches based on probabilistic graphical models \cite{hawe2019inferring,koller2009probabilistic}. {We note that several of these methods were originally developed for transcriptomic datasets and are thus discussed in Section \ref{sect:network_inference_comparison}.} In probabilistic graphical models, nodes are modeled as random variables and edges correspond to statistical dependencies \cite{koller2009probabilistic},  where each data modality is represented as a different node type \cite{Chen2014selection,Sedgewick2018mixed}. {A key modeling challenge when handling multiple types of measurements is to specify the appropriate probability distributions for each data modality \cite{Chen2014selection,Sedgewick2018mixed}. Furthermore, the larger number of variables of multimodal data introduces additional scalability issues for learning the structure of probabilistic graphical models such as general Bayesian networks. Several heuristics such as focusing on promising parents \cite{friedman2013learning, schmidt2007}, exploiting modularity of molecular networks \cite{segal2005learning}, or approximating joint probability distributions as done in dependency networks \cite{greenfield2013robust, heckerman2000dependency,roy2013integrated} have enabled these models to scale to thousands of variables.} 

Once the networks have been defined, they can be further clustered into modules to identify potential functional groupings among the nodes \cite{choobdar2019assessment,mitra2013integrative,newman2006modularity}. Unsupervised learning of graph structure from multi-omic data lends itself naturally to the inference of gene regulatory networks \cite{baur2020data}, where node types represent target genes and protein regulators. Protein regulators can be further modeled based on their observed mRNA levels or their hidden activity levels \cite{miraldi2019leveraging}. While such approaches do not need any edge-level information, if any, potentially noisy, information is available, this can be incorporated as a graph prior to guide the structure learning \cite{greenfield2013robust,miraldi2019leveraging,siahpirani2017prior}. 

The availability of single-cell multi-omic datasets has also opened up challenges  that can be tackled with heterogeneous network inference \cite{demetci2022scot,heumos2023best}. One such problem is to infer cell-cell networks with nodes corresponding to cells, node types corresponding to different modalities (e.g., scRNA-seq, scATAC-seq) or time points (or both), and edges representing different semantics such as similarity or lineage relationships. Due to the size and sparsity in these data, dimensionality reduction is typically performed prior to inference of network structure. Non-negative matrix factorization, independent components analysis, and variational autoencoders are common dimensionality reduction approaches for single-cell multi-omic datasets. After dimensionality reduction, graph learning can be done using the $k$-nearest neighbor approach \cite{ butler2018integrating} or with optimal transport \cite{demetci2022scot,schiebinger2019optimal}. Graphs based on $k$-nearest neighbors, with different distance measures, are straightforward to implement and frequently used in practice, while optimal transport’s framework to match probability distributions of cells can be used to capture fine-grained cell dynamics. 

\xhdr{Pathway reconstruction for interpretation of multi-omic data}
Heterogeneous networks offer a powerful framework to integrate, interpret, and reconcile missing and noisy measurements commonly seen in multi-omic experiments  \cite{haque2017practical,peck2021boosting}. The task of pathway reconstruction takes as input multi-omic measurements of different biomolecules represented as node types and a large background molecular network. It outputs a sparse subnetwork with high-quality connections among the relevant biomolecules \cite{garrido2022integrating} (Fig. \ref{fig:multimodal_fig1}C). The background networks typically contain PPIs and may also include protein-DNA, protein-RNA, or protein-metabolite interactions to match the available -omic data. Paths from one relevant biomolecule to another in the background network can help prune irrelevant biomolecules and identify those that may play critical roles in the overall biological process but were missed in the -omic measurements \cite{paull2013discovering,pirhaji2016revealing,tuncbag2016network,winkler2022novo}. {Note that this task also relates to condition-specific network inference discussed in Section \ref{sect:network_inference_comparison} and multi-omic module discovery discussed in Section \ref{sect:personalized_medicine} for discovery of dysregulated pathways in diseases such as cancer.}

The sparse subnetwork obtained depends on the choice of optimization algorithm and its parameters. Some pathway reconstruction algorithms are computationally efficient, based on shortest paths \cite{ritz2016pathways} or network flow \cite{yeger2009bridging}. Despite their algorithmic simplicity, these methods can still effectively prioritize biologically relevant nodes and interactions. Network flow-based {methods} can scale across multiple experiments by relying on the multicommodity flow approach, which identifies nodes and edges that are unique and shared across conditions \cite{gosline2012samnet}. General integer linear programming approaches \cite{chasman2014inferring, ourfali2007spine} support arbitrary node, edge, and path constraints. These provide the greatest customization for a particular multi-omic dataset but less scalability and reusability across applications. Intermediate approaches such as the Prize-Collecting Steiner Forest \cite{tuncbag2013simultaneous} are computationally difficult to solve exactly but can be approximated efficiently. For instance, the Omics Integrator software \cite{tuncbag2016network} based on the Prize-Collecting Steiner Forest algorithm adds prizes to nodes that should be included in the sparse subnetwork and costs to edges based on their reliability. {Omics Integrator also includes a module to estimate prizes for active TFs based on chromatin accessibility, gene expression, and DNA-binding motifs.} Its parameters control the tradeoff between node prizes and edge costs, a penalty for including nodes with high degree, and a penalty for the number of connected components in the subnetwork.
      
Heterogeneous pathway reconstruction is especially powerful because network connections between different types of biomolecules can be combined to reveal more complete and explanatory pathways. For instance, a TF that activates differentially expressed genes detected with RNA-seq may be inferred to be regulated by an upstream phosphorylated kinase detected with mass spectrometry. A study of Kaposi's Sarcoma-associated Herpesvirus infection \cite{sychev2017integrated} illustrates the data types and algorithms involved, and biological insights gained in multi-omic pathway reconstruction. The authors profiled the proteomic and phosphoproteomic changes in endothelial cells induced by viral infection using mass spectrometry and gene expression changes with RNA-seq. They used TF binding motifs and a statistical enrichment test with the gene expression data to identify potentially relevant transcriptional regulators. Then, they applied Omics Integrator \cite{tuncbag2016network} to combine the transcriptional regulators, proteomic changes, phosphoproteomic changes, and a PPI background network in order to obtain a holistic view of the endothelial cell response to infection. Ultimately, this analysis revealed peroxisome-related proteins to be an important part of the response. This network-based insight was supported with follow-up wet laboratory experiments \cite{sychev2017integrated}.

\xhdr{Network alignment} 
In network biology, network alignment has traditionally been used to compare species' PPI networks \cite{Emmert-Streib2016,faisal2015post,Guzzi2017,ma2022heuristics,Sharan2006,PNAMNA}. In this context, network alignment aims to find a node (protein) mapping between the compared networks that uncover regions of high topological (and often sequence) conservation, with the hypothesis that the resulting aligned nodes and network regions are evolutionary conserved or functionally similar. Finding such a node mapping is closely related to the NP-complete subgraph isomorphism problem, making the network alignment problem NP-hard \cite{faisal2015post}. 

Even when comparing PPI networks, which are homogeneous, network alignment can be viewed as a multimodal data integration task. This is because an alignment (i.e., node mapping) in a ``composed view''  results in a heterogeneous (specifically, multiplex) network whose ``supernodes'' contain mapped nodes from the individual homogeneous networks and whose edges are of distinct types, indicating which one(s) of the compared networks the given edge is present in under the given node mapping (Fig. \ref{fig:multimodal_fig1}D). More recently, approaches have been proposed for aligning heterogeneous networks in biology \cite{gu2018HNA,LHetNetAligner} and other domains \cite{chen_fascinate_2016,yan_dissecting_2022}. Below, we discuss algorithmic principles of traditional alignment of homogeneous networks and then comment on the alignment of heterogeneous networks.

Analogous to sequence alignment, alignment of homogeneous networks can be local or global \cite{Meng2016}. Both have (dis)advantages \cite{Guzzi2017}. Also, network alignment can be pairwise (between exactly two networks) or multiple  (between more than two networks) \cite{multiMAGNA++}. The latter has traditionally been expected to lead to deeper biological insights as it aligns all considered networks simultaneously as opposed to one pair at a time; however, a recent evaluation showed that this is not always the case \cite{PNAMNA}. At the same time, multiple network alignment is computationally more complex \cite{multiMAGNA++}.

Network alignment has two main algorithmic components \cite{faisal2014global}. First, topological similarity between nodes across the compared networks is computed via some measure of node conservation; graphlet-based measures (Section \ref{sect:higher_order_network_analysis}) are among state-of-the-art \cite{gu2018HNA,newaz2019bookchapter}. Second, an alignment strategy quickly identifies alignments that optimize some objective function accounting for total node and ideally also edge conservation under the given node mapping. That is, a good alignment should both map similar nodes to each other and conserve many edges. Original alignment strategies were of the seed-and-extend type \cite{GRAAL,singh2008global,WAVE}. The extension around highly similar ``seed'' nodes, by adding mapped nodes incrementally to build the alignment one step at a time, is intended to explicitly improve node conservation of the resulting alignment, but edge conservation only implicitly. To improve edge conservation explicitly as the alignment is constructed, rather than only evaluating it after the fact, another type of alignment strategy -- a search algorithm -- was introduced. Here, entire alignments are explored, and the one that scores the highest based on the given (e.g., edge conservation-based) objective function is returned, using, e.g., genetic algorithms \cite{MAGNA,multiMAGNA++,DynaMAGNA++,MAGNA++} or simulated annealing \cite{SANA}.

A recent algorithmic shift in network alignment has been from unsupervised to supervised, data-driven alignment \cite{TARA,TARA++}. Traditional network alignment uses the notion of topological similarity to quantify how close to isomorphic two nodes' extended  network neighborhoods are. A major issue is that regardless of the considered similarity measure, aligned nodes often do not correspond to nodes that should actually be mapped, i.e., that are functionally related \cite{TARA}. Specifically, when comparing species' PPI networks, aligned nodes do not correspond to proteins that are involved in same biological processes. This is why a move was made from optimizing topological similarity to learning from the data what kind of topological relatedness corresponds to functional relatedness, without assuming that topological relatedness means topological similarity \cite{TARA}. For example, topological similarity will aim to match a triangle in one network to a triangle in another network, and a square in the former to a square in the latter. Yet, due to biological variation or noise in PPI data, perhaps it is the triangle in the first network that is functionally related and should thus be matched to the square rather than the triangle in the second network, which is what topological relatedness would aim to learn from the data.  This resulted in moving from traditional unsupervised alignment (functional labels of nodes, e.g., biological processes of proteins in PPI networks, being used to evaluate an alignment only after it is produced) to supervised, data-driven alignment (functional labels of nodes being used during the process of constructing an alignment, to learn patterns of topological relatedness). A pioneering data-driven network alignment method used traditional machine learning, i.e., user-predefined (graphlet-based) features \cite{TARA,TARA++} and standard classifiers (e.g., logistic regression), while a follow-up effort used deep learning and specifically GNNs \cite{ding2023supervised}. 

Finally, going back to alignment of heterogeneous networks, an earlier attempt in biology was still to align homogeneous networks to each other, where the heterogeneity came from the fact that the individual homogeneous networks being compared were of different types: one was a human PPI network whose nodes were proteins, and the other was a disease-disease association network whose nodes were diseases \cite{Wu2009}. Then, the goal of aligning the two networks was to identify causative genes/proteins and their pathways underlying disease families. But, because each of the compared networks was homogeneous, a homogeneous network alignment approach sufficed for their comparison. A more recent effort towards actually aligning one heterogeneous network to another, each with different node and edge types (or colors), was extending the existing notions of homogeneous graphlet-based node similarity/conservation as well as homogeneous edge conservation (discussed above) into their heterogeneous (or colored) counterparts, and then extending the existing seed-and-extend or search alignment strategies (discussed above) to find high-scoring alignments with respect to the new heterogeneous conservation measures \cite{gu2018HNA}. In evaluations on synthetic and real biological networks, the heterogeneous  methods led to higher-quality alignments and better robustness to noise in the data than their homogeneous counterparts \cite{gu2018HNA}. Two types of heterogeneous biological networks were considered: first, PPI networks were aligned to each other, where nodes (proteins) were colored according to whether they were involved in aging, cancer, and/or Alzheimer's disease; second, protein-GO term networks were aligned to each other, where such a network had two types of nodes -- proteins and GO terms -- and three types of edges -- PPIs, protein-GO term annotations, and GO term-GO term semantic similarity associations \cite{gu2018HNA}. This effort \cite{gu2018HNA} aligned heterogeneous networks globally. In parallel, an approach for their local alignment was proposed \cite{LHetNetAligner}.

{Ideas from machine learning-based embedding of heterogeneous networks (Section \ref{sect:networkml-intro}) in biology \cite{pio2021multiverse} and other domains \cite{wang2022multiplex,wang2022survey} could be extended to heterogeneous network alignment. However, to our knowledge, such extension has not yet been carried out in biology but it has been carried out in other domains such as social, information, or technological networks \cite{cai2023resolving,wang2022network,xiong2021contrastive,zhang2020nettrans,zhang2019origin,zheng2018heterogeneous}. Note that in \cite{zhang2020nettrans,zhang2019origin},} the heterogeneity of considered networks came from node/edge attributes rather than explicit node/edge types. In these two studies, GNNs were used to first find an embedding of nodes of the compared networks, and then the network alignment problem was viewed as a point registration problem \cite{zhang2019origin} or a neural network transformation problem \cite{zhang2020nettrans}.

\xhdr{Inference of and reasoning on biomedical knowledge graphs}
Biomedical knowledge graphs (BKGs), which describe semantic relationships between biomedical entities, are among the richest examples of heterogeneous networks \cite{nicholson2020constructing}. BKGs aim to combine facts about diverse biomedical entities, which can range from genes to individual patients as well as measurements associated with them. BKGs represent biological facts using ``subject-predicate-object'' triples as the fundamental unit, with the subject and object corresponding to nodes in the graph and the predicate (also called a relation) corresponding to a directed edge, possibly of different types, between the nodes. For example, Chlorin e6-PDT (subject) reduced (predicate) cell proliferation (object); Fig. \ref{fig:multimodal_fig1}E. Exemplar active BKG projects, each taking a unique approach, include Scalable Precision Medicine Knowledge Engine (SPOKE)\footnote{\url{https://spoke.rbvi.ucsf.edu}}  \cite{morris2010ceres}, BioThings Explorer\footnote{\url{https://explorer.biothings.io}} \cite{fecho2022progress,lelong2022biothings}, biomedical ``corner'' of Wikidata\footnote{\url{https://www.wikidata.org}}  \cite{manske2019genedb,page2022wikidata,waagmeester2020wikidata}, and PrimeKG \cite{chandak2023building}.

BKGs have emerged as powerful frameworks for diverse biomedical applications \cite{nicholson2020constructing} including drug repurposing (e.g., Hetionet \cite{himmelstein2017systematic} and SPOKE \cite{morris2010ceres}), rare disease diagnosis \cite{alsentzer2022deep}, and biomarker discovery (e.g., SPOKE \cite{himmelstein2015heterogeneous}). BKGs leverage graph databases like Neo4j and Virtuoso, and semantic web standards like the Resource Description Framework for their backend. BKGs leverage over a hundred years of graph theory to enable operations on first neighbors, paths, centralities, and other network components, as well as semantics, inference, and reasoning. There are a number of computational challenges that emerge to maximally extract the information encoded in BKGs for diverse biomedical applications ranging from construction of BKGs to reasoning with BKGs \cite{peng2023knowledge}. For example, advanced, multi-hop queries specifying node and edge types are essential to navigating heterogeneous network representations of biomedical knowledge{; ``multi-hop'' refers to having to traverse at least two edges in the graph}. Many of these challenges have been approached using similar methods of network inference as previously described (e.g., link prediction) as well as more recently with graph representation learning approaches discussed in Section \ref{sect:networkml-intro}. 

Equally important is the question of representation of biomedical and biological literature to enable advanced queries and reasoning. Traditional BKGs assume that all knowledge can be represented as subject-predicate-object tuples and are constructed using tuple extraction techniques based on machine learning. A simple postprocessing algorithm can extract the tuples from any sentence and represent them as links between nodes on the BKGs. However, traditional BKGs have ignored the conditions (e.g., patient age or environment) of the facts, which capture essential contexts for knowledge exploration and inference. Recently, a new type of BKG, Condition-aware BKG (CondBKG  \cite{jiang2020biomedical}), has been introduced, which considers both facts and their conditions in the biomedical statements. Unlike traditional BKGs which have only one layer of subject-predicate-object tuples, CondBKG is a three-layered information-lossless representation of BKGs. The first layer has biomedical concept and attribute nodes; the second layer represents both biomedical fact and condition tuples by nodes of the predicate phrases, connecting to the subjects and objects in the first layer; the third layer has nodes that represent statement sentences as their textual attributes and connect to fact and/or condition tuples in the second layer (Fig. \ref{fig:multimodal_fig1}E). CondBKG is constructed from a machine learning model's output tuples. 
Given a statement sentence and its context (e.g., nearby sentences) in a scientific article, the model learns from multiple types of input signals of sentence (e.g., word embeddings and part of speech tags) and predicts one or multiple tuples. CondBKG has 18.1 million fact tuples, 7.5 million condition tuples, 10.9 million concept nodes, and 703 thousand attribute nodes. CondBKG preserves more knowledge from unstructured text than traditional flat BKGs and can be used to answer tailored queries, such as what factors increase or reduce cell proliferation and their conditions (Fig. \ref{fig:multimodal_fig1}E). CondBKG can provide a good understanding of biomedical and biological statements and supports diverse applications for biomedical knowledge discovery.

\xhdr{Network-of-networks analysis}
Biological systems function at different scales of organization. Thus, network-of-networks analysis (Fig. \ref{fig:multimodal_fig1}A) is an exciting, still relatively unexplored area of research. This topic has received an increasing amount of attention only in recent years. This is likely because it has been increasingly recognized that network-of-networks representations of various biological data can be obtained: {(1)} given that different diseases tend to manifest in different tissues, nodes (diseases) in a disease similarity network can be represented as their associated tissue-specific PPI networks \cite{ni2016disease}; {(2)}  nodes  in a PPI network can be represented as protein structure networks \cite{gao2023hierarchical,gu2022modeling}; {(3)} nodes  in a network of interacting molecules  can be represented as molecular graphs \cite{wang2022powerful,wang2020gognn}; {(4)} nodes in a bipartite graph containing interactions between drugs and their target proteins can be represented as drug molecule graphs and target protein structure networks, respectively \cite{chu2022hierarchical}. Note that not all existing network-of-networks studies originate in the biology domain. Some have been proposed and evaluated in other domains, such as on text and social network datasets \cite{li2022semi}. 

The studies that have analyzed biological network-of-networks data typically perform different network analysis and application tasks, as follows. The task of node ranking was applied to candidate disease gene prioritization {from the network-of-networks of type (1) above} \cite{ni2016disease}. The task of link prediction was applied to predicting interactions between proteins {from the network-of-networks of type (2) above} \cite{gao2023hierarchical}, between molecules such as drugs {from the network-of-networks of type (3) above} \cite{wang2022powerful,wang2020gognn}, or between drugs and their target proteins {from the network-of-networks of type (4) above} \cite{chu2022hierarchical}. A new task was introduced -- that of entity label prediction -- which merges the two traditionally isolated tasks of node (protein) classification at the higher scale containing a PPI network and graph (also protein) classification at the lower-scale containing protein structure networks \cite{gu2022modeling}. This task was applied to {predicting protein functions from the network-of-networks of type (2) above}  \cite{gu2022modeling}. Given that the different approaches were proposed for different tasks/applications, they have typically not been evaluated against each other. It remains unclear whether the different approaches can be effectively used in tasks/applications other than those they were proposed for, as well as what (dis)advantages of each approach are on the methodological level. With the increasing availability of network-of-networks data and the increasing number of approaches for network-of-networks analysis, the need for proper method evaluation will only continue to gain importance. This will require all studies to make their data and code publicly available and easy to use. According to our exploration of the existing network-of-networks studies discussed above, this is not always true.

%% file: 040.higher_order_rtf.tex
\xhdr{Need for higher-order graph representations of biological systems} 
This paper, unless explicitly noted otherwise, deals with traditional pairwise graphs (or simply graphs).  Such a graph represents the organization of a biological system as a network of pairwise interactions between biomolecules (e.g., a PPI is represented as an edge connecting two proteins, and a transcriptional regulatory interaction is represented as a directed edge from a TF to a gene).
However, these interactions often involve additional components and the interactions themselves can be regulated by other components~\cite{battiston2020networks}. In other words, there is often a need to capture interactions between multiple (two or more) nodes rather than between exactly two nodes (as is the case with pairwise graphs). Several higher-order graph ideas have been proposed in the literature to overcome the limitations of pairwise graphs. There are two general categories of such ideas. 

The first category still works with pairwise graphs but relies on either higher-order dependencies between two nodes  \cite{xu2016representing} or small subgraphs \cite{newaz2019bookchapter}, as follows. Regarding higher-order dependencies, it was shown that when representing sequential data such as global shipping traffic as networks, assuming the first-order dependency, i.e., that the next movement of traffic depends only on the current node, and thus discounting the fact that the movement may depend on several previous steps, can yield inaccurate network analysis results \cite{xu2016representing}. This is because data derived from many complex systems can show up to fifth-order dependencies between two nodes. Consequently, an approach was proposed for capturing variable orders of dependencies between pairs of nodes \cite{xu2016representing}. Regarding subgraphs, these can be viewed as ``higher-order coordinated patterns'' between two or more nodes of a pairwise graph \cite{battiston2020networks}; a subgraph captures first-order dependencies (as discussed above and defined in \cite{xu2016representing}) between multiple nodes in a pairwise graph. Examples of subgraph types are cycles (e.g., a triangle or a square) or cliques (the densest of all subgraph types, containing all possible edges between their nodes) \cite{battiston2020networks}. Two general categories of subgraphs exist: graphlets \cite{Przulj2007} and network motifs \cite{Milo04}. Two key differences exist between them: graphlets are induced subgraphs while network motifs are not, and network motifs need to be statistically significantly over-represented in a pairwise graph compared to a null (i.e., random graph) model while graphlets do not rely on a null  model.

Both higher-order dependencies and subgraphs in pairwise graphs from the first category fail to directly account for interactions between more than two nodes in a network. An alternative is the second category of higher-order graph ideas -- to explicitly consider higher-order graph structures. Here, while simplicial complexes are a theoretic possibility, they have assumptions that are practically too strong in some  systems \cite{battiston2020networks}. The next most general idea of higher-order interactions that is at the same time less constraining and thus more practical are hypergraphs \cite{battiston2020networks}.

{Higher-order dependencies (as discussed above and defined in \cite{xu2016representing}) have not yet received attention in the biology domain, which is why we do not discuss this idea further.} Graphlets in pairwise graphs (or simply graphlets), hypergraphs, and graphlets in hypergraphs (i.e., hypergraphlets) have received significant attention in the biology domain, which is why the following sections discuss these topics in more detail. While network motifs have also received attention, it remains unclear which random graph model fits real-world networks the best and should thus be used for network motif identification \cite{ArtzyRandrup04,newaz2019bookchapter}, which is why we do not discuss network motifs further. 

\xhdr{Graphlets} 
Graphlets, small subgraphs, are Lego-like building blocks of a network. More formally, they are connected, non-isomorphic, induced subgraphs of a graph \cite{Przulj2004}. Because counting of large graphlets in a large network is time-consuming, in practice, graphlets on up to five nodes have typically been studied. Graphlets were originally proposed as subgraphs of undirected, homogeneous, static, unordered, and pairwise graphs \cite{newaz2019bookchapter}. More recently, they were extended to their directed \cite{Lugo2014,Sarajlic2016}, heterogeneous \cite{gu2018HNA}, dynamic \cite{hulovatyy2015exploring}, ordered \cite{GRAFENE,GRALIGN}, or hypergraph \cite{Gaudelet2017HG,Lugo-Martinez2021} counterparts, respectively; the latter are called hypergraphlets and are discussed more below after hypergraphs are introduced. The following concepts are discussed for original graphlets, but they generalize to the more data-rich counterparts as well. 

In a graphlet, nodes can correspond to different symmetry groups called automorphism orbits (or just orbits for simplicity) \cite{Przulj2007}. For example, in a graphlet corresponding to the 3-node path (e.g., $a-b-c$), the two outer nodes ($a$ and $c$ in our illustration) are symmetric to each other and thus belong to the same  orbit, while the middle node ($b$) is in its own orbit. As another example, in a clique, all nodes are symmetric to each other and thus belong to the same  orbit. There are 15 orbits for 2-4-node graphlets and 73 for 2-5-node graphlets. This concept of graphlet orbits can be used to quantify a node's extended network neighborhood into a 15- or 73-dimensional embedding, often called the node's graphlet degree vector (GDV) \cite{milenkovic2008uncovering}. This vector counts how many times a node of interest touches (or participates in) each of the considered graphlets at each of their orbits. By computing GDV for each node in a network, one can obtain the network's GDV matrix, whose entry $(i,j)$ contains the information of how many times node $i$ touches orbit $j$ \cite{milenkovic2008uncovering,newaz2019bookchapter}. Note that there exist an analogous concept of edge (rather than node)  as well as node pair orbits, GDVs, and GDV matrices \cite{Hulovatyy2014,Solava2012}.

GDV matrices of networks have been used as features to compare extended  neighborhoods of nodes (edges, node pairs) in the same network, extended  neighborhoods of nodes (edges, node pairs) across different networks, or structures of entire networks \cite{newaz2019bookchapter}. These, in turn, have been used in numerous computational tasks, such as network alignment, alignment-free network comparison, graph classification, node classification, network de-noising via link prediction, inference of a condition-specific network or pathway reconstruction, network clustering, and node centrality computation, as well as for various application problems, such as studying human aging, protein folding and function, cancer and other diseases, pathogenicity, or mental health (e.g., depression and anxiety), as briefly discussed in other sections {\cite{arici2023unveiling,Liu2020,liu2021heterogeneous,magnano2021automating,newaz2019bookchapter,newaz2022multi,Solava2012}}.

\xhdr{Hypergraphs} 
Hypergraphs provide powerful representations by generalizing edges between exactly two nodes to hyperedges that involve multiple nodes~\cite{Berge1973}. For example, protein complexes, which involve simultaneous interactions among multiple proteins that carry out function only as a group, are effectively represented using undirected hypergraphs, where each node is a protein and each undirected hyperedge (a set of nodes) is a complex~\cite{klamt2009hypergraphs}. Under this representation, complexes that share interactors can be disambiguated, thus allowing more flexibility to capture multiple functionalities on the same set of nodes. Signaling pathways, on the other hand, are represented using directed hypergraphs in which proteins are represented by nodes and reactions are represented by directed hyperedges~\cite{ritz2014signaling}.

\begin{figure*}[t!]
  \centering
    \includegraphics[width=0.79\linewidth, trim=0cm 0.5cm 0cm 0cm]{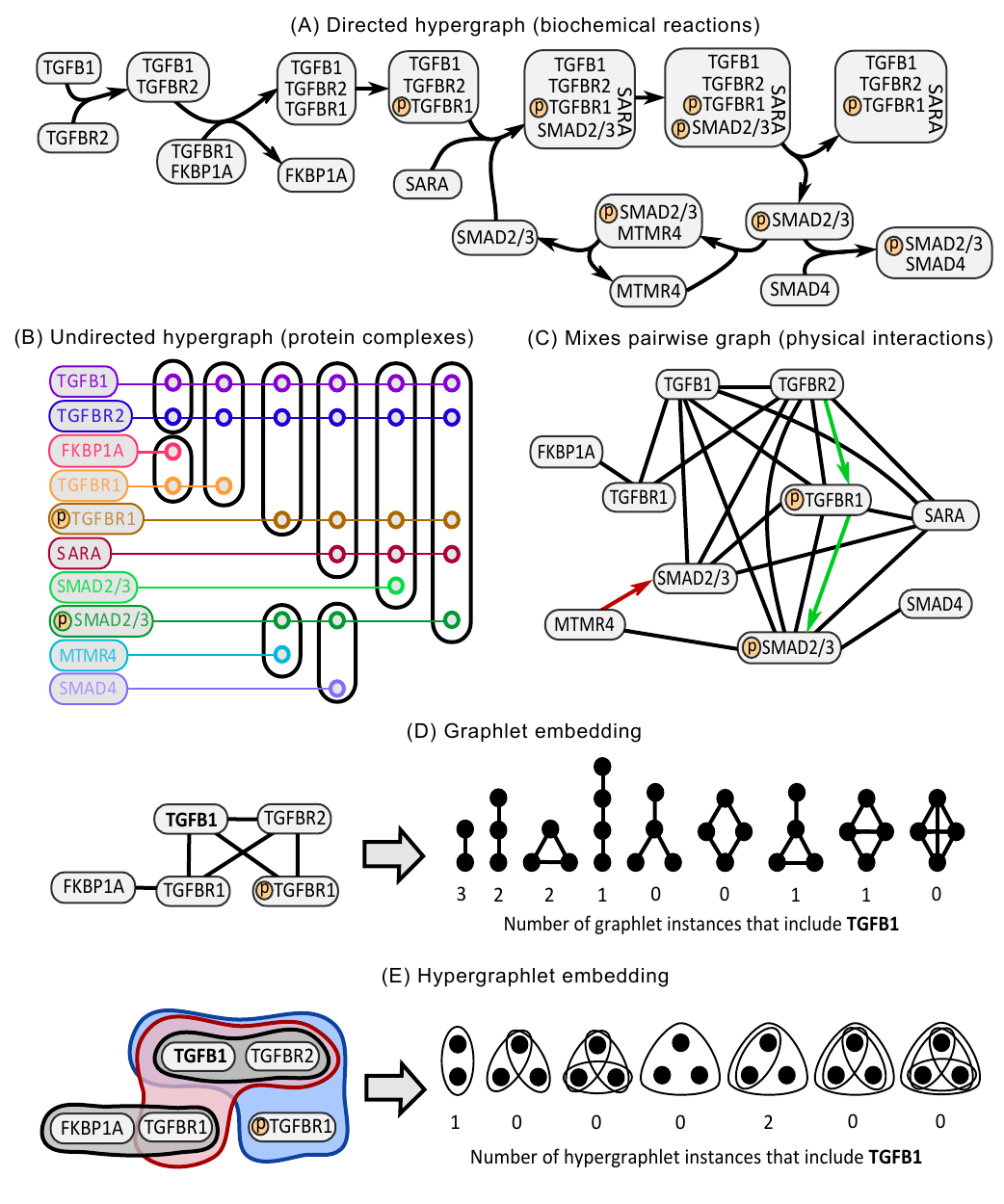}
     \caption{Graph representations of nine reactions from Reactome's TGF$\beta$ signaling pathway. 
     \textbf{(A)} In a directed hypergraph, each hyperedge captures a reaction (``p'' denotes phosphorylation). \textbf{(B)} In an  undirected hypergraph, each hyperedge captures a protein complex. \textbf{(C)} In a (mixed) pairwise graph, each edge captures a  pairwise interaction. ``Mixed'' refers to having both directed and undirected edges in the graph. Black undirected edges denote physical interactions; green directed edges denote phosphorylation; red directed edges denote dephosphorylation. \textbf{(D)} A node in a pairwise graph can be represented as a vector of graphlet counts. The number of 2-, 3-, and 4-node graphlet instances that include TGFB1 in the graph on the left are shown. \textbf{(E)} A node in an undirected hypergraph can be represented as a vector of hypergraphlet counts. The number of 2- and 3-node hypergraphlet instances that include TGFB1 in the hypergraph on the left are shown. In panels (D)-(E), only the (hyper)graphlet-level counts are shown for simplicity, i.e., (hyper)graphlet orbits are not shown nor considered when doing the counting. However, in practice, the more detailed orbit-level counts are computed rather than the (hyper)graphlet-level counts. }
     \label{fig:TGFb}
\end{figure*}

Fig.~\ref{fig:TGFb} shows an example of nine reactions from the transforming growth factor-beta (TGF$\beta$) signaling pathway~\cite{gillespie2022reactome} and their representation using higher-order graph frameworks. In this example, TGF$\beta$1 binds to the TFG$\beta$ receptor and phosphorylates SMAD2/3, which in turn binds to SMAD4; SMAD2/3 are subsequently dephosphorylated by MTMR4. The signaling reactions are captured by a directed hypergraph with nine  hyperedges connecting proteins (which may be phosphorylated) and protein complexes (Fig.~\ref{fig:TGFb}A). Without the directed hyperedges, we have a series of overlapping protein complexes, the structure of which provides some insights into how the protein complexes form (Fig.~\ref{fig:TGFb}B). Directed and undirected hypergraphs offer more information than a graph that only captures pairwise physical interactions in this cascade (Fig.~\ref{fig:TGFb}C). If dealing with the pairwise graph representation in Fig.~\ref{fig:TGFb}C, graphlets can help characterize the local topology of a specific node (Fig.~\ref{fig:TGFb}D) or an entire network, as discussed above. If dealing with the hypergraph representation from Fig.~\ref{fig:TGFb}A-B, hypergraphlets, discussed below, can be used to quantify topology (Fig.~\ref{fig:TGFb}E).

A shortcoming of pairwise graphs in representing multi-component interactions is that some paths may be lost~\cite{murgas2022hypergraph} or ghost paths can be  created~\cite{pandey2007functional} while contracting a multi-way interaction into a set of pairwise interactions.  For example,  as seen in Fig.~\ref{fig:TGFb}A, the interaction between TGF$\beta$1 and SMAD2/3 occurs when TGF$\beta$1 is part of the TGF$\beta$ complex that is phosphorylated, but this information is lost in the  pairwise graph representation shown in Fig.~\ref{fig:TGFb}C. In addition,  contracting multi-way interactions into pairwise interactions results in the replication of interactions between multiple components, inflating subgraph density, multiplicity of paths, and node degrees; while also shortening paths. Generalization of notions such as density or centrality to hypergraphs can therefore provide more reliable insights into the topology and dynamics of biological networks~\cite{feng2021hypergraph}.

In addition to reducing representation loss, hypergraphs also offer meaningful algorithmic advantages. Owing to the graph duality property where each graph can be represented as a hypergraph by inverting nodes and edges of the original graph into hyperedges and nodes, respectively, of a dual graph, hypergraph representations offer a possibility to unify methodology. For example, node classification, edge classification, and link prediction on pairwise graphs can all be seen as node classification on (extended) dual hypergraphs \cite{Lugo-Martinez2021}. This allows for the development of general methodologies and software that could support statistical inference tasks on biological networks.

To date, the application of hypergraphs in biological network analysis is limited because of constraints posed by the availability of data and annotations (or lack thereof). In cellular signaling, post-translational modifications play a central role in multi-way interactions among cellular components, yet only a small fraction of post-translational modifications are well-characterized~\cite{needham2019illuminating}. As biotechnology advances and more data are generated, the availability of algorithms that solve fundamental problems on hypergraph representations, therefore, has the potential to guide data generation and curation of annotations. 

\xhdr{Hypergraph algorithms} 
In the broader computer science community, hypergraph algorithms exist for several problems including shortest paths, random walks, and  clustering~\cite{ausiello2017directed,cambini1997flows,ducournau2014random,gao2014dynamic,zhou2006learning}. Within the context of network biology, hypergraphs have been used to study metabolic networks~\cite{klamt2009hypergraphs}, clusters in PPI networks~\cite{ramadan2004hypergraph}, and shortest paths in signaling pathways. This final application is the best developed use of directed hypergraphs in network biology. Hence, we focus our discussion on it. 

Defining reachability in directed hypergraphs is significantly more complex than in pairwise graphs. A key principle is that the nodes in the head of a hyperedge are reachable from some source only if \emph{all} the nodes in the tail are themselves reachable from that source. This principle expresses the natural concept that for any product of a reaction to form, all the reactants must be present. The notion of B-reachability formalizes this idea~\cite{ausiello2017directed,ritz2014signaling}. The challenge now is that computing B-hyperpath with the smallest number of edges is an NP-complete problem, even when the tail and head of each hyperedge contain at most two nodes and we are interested only in acyclic  hyperpaths~\cite{ritz2014signaling}. An initial approach proposed a mixed-integer linear program to compute optimal hyperpaths~\cite{ritz2014signaling}, applying it with success to the Wnt signaling pathway in the NCI Pathway Interaction Database. In practice, a drawback of this method was that a very large number of nodes without any incoming hyperedge had to be included among the sources for any meaningful hyperpath to exist. A later technique relaxed the definition of B-hyperpath~\cite{franzese2019hypergraph} to address this problem. As another alternative, an efficient heuristic approach can handle cyclic hyperpaths and computes optimal ones in practice ~\cite{krieger2022heuristic}. An exact cutting-plane algorithm can also compute shortest hyperpaths with  cycles while being efficient in practice on both the NCI Pathway Interaction Database and Reactome~\cite{krieger2023computing}. Finally, similar problems have been studied in the context of metabolic networks. Here, the notion of shortest path is generalized to a factory, which also takes reaction stoichiometry into account. A mixed-integer linear program can find factories with the fewest reactions and {accommodate} 
{negative} regulation~\cite{krieger2022computing}.

\xhdr{Statistical learning on hypergraphs} 
Hypergraphs can be approximated by pairwise graphs (e.g., star expansion, clique expansion \cite{Agarwal2006}), but such approximations do not retain all properties of the original hypergraphs (e.g., the cut properties \cite{Ihler1993}). Therefore, methods directly developed for learning on hypergraph data can offer practical advantages. A number of such approaches have emerged \cite{Antelmi2023,chitra2019random,Cong1991, Leordeanu2012, Lugo-Martinez2021, Maleki2022,Wachman2007}; however, accurate learning on hypergraphs is often hindered by NP-hardness issues \cite{Gartner2003, Hein2013, Purkait2017} and, thus, methods developed to directly deal with hypergraph data often trade accuracy for scalability.

A common theme in statistical learning on hypergraphs is finding a typically high-dimensional representation, or an embedding, of the data, and subsequently applying traditional machine learning to learn some concept; see Section \ref{sect:networkml-intro} for more details. These methods can work at the level of entire graphs for graph classification, or at the level of nodes (edges), for node (edge) classification and link prediction. A well-known graph classification problem is the prediction of toxicity of chemical molecules \cite{Vishwanathan2010}, where the nodes are atoms, and the edges are bonds, both of different types, or prediction of protein function \cite{Borgwardt2005}. Examples of popular node/edge classification problems are function prediction for proteins/protein complexes in PPI networks or for amino acid residues in protein structure networks \cite{Lugo-Martinez2016,Vacic2010}. An example of a link prediction problem is the task of de-noising and completion of the PPI network itself, as also discussed in Section \ref{sect:network_inference_comparison}. 

{Embeddings} are often formalized via kernel-based approaches or representation learning (Section \ref{sect:networkml-intro}), thus allowing the practitioners to use both finite- and infinite-dimensional representations. Well-performing kernel approaches (kernels are symmetric, positive semi-definite similarity functions defined on pairs of objects, that allow efficient learning \cite{Shawe2004}) include random walks \cite{Wachman2007} and hypergraphlet counting \cite{Lugo-Martinez2021}. Hypergraphlets are typically defined as small, connected, (rooted) hypergraphs, often with a finite number of node and edge types \cite{Lugo-Martinez2021}. They are a non-trivial extension of (pairwise) graphlets discussed above \cite{Lugo2014, milenkovic2008uncovering, Przulj2007,Przulj2004,Shervashidze2009, Vacic2010}, with both illustrated in Fig. \ref{fig:TGFb}D-E. As with graphlets, the appeal for counting hypergraphlets derives from the graph reconstruction conjecture \cite{Bondy1977}. Though proved only for certain types of graphs (e.g., trees), the graph reconstruction conjecture postulates that a large graph of size $n$ can be reconstructed up to isomorphism from the counts of all subgraphs up to the size of $n-1$. A stronger version of the conjecture allows for such reconstruction for subgraphs up to the size of some $k<n-1$. Under these conditions, hypergraphlet counting approaches can lead to embeddings that allow universal approximation on hypergraph data. Another approach, relying on neural-network graph embeddings, allows for scaling hypergraph-based approaches to very large graphs \cite{Maleki2022}.

{Additional approaches for hypergraphs exist, which are based on deep learning~\cite{gui2016large,tu2018structural}. Among these, a} prominent example utilizes a GNN based on self-attention to effectively learn embeddings of the nodes and predict hyperedges for non-$k$-uniform heterogeneous hypergraphs, enhancing the generalizability~\cite{zhang2020hypersagnn}. This approach and its extensions have been applied to studying chromatin biology~\cite{zhang2020matcha,zhang2022multiscale} and predicting genetic interactions for a group of genes, specifically trigenic interactions, thereby significantly expanding the quantitative characterization of higher-order interactions~\cite{zhang2020dango}. 

\xhdr{Limitations} 
Three major issues confront the wide adoption of hypergraph-based representations in network biology. Databases such as Reactome \cite{gillespie2022reactome} contain well-curated reaction networks that are amenable to representations as generalizations of directed hypergraphs. The first issue is that these resources remain incomplete and rely on manual curation. One promising direction of research is to analyze pairwise graphs to automatically infer reactions. An elegant example is an approach that uses properties of chordal graphs to convert a graph representation of a signaling pathway as a nested tree of protein complexes~\cite{zotenko2006decomposition}. A graph is chordal if every pair of nodes in every cycle of length four or more is connected by an edge. Since PPI networks are not necessarily chordal, the authors augment them with additional edges, e.g., those that connect weak siblings, i.e., pairs of nodes that have identical neighbor sets but are themselves not connected by an edge. If the resulting graph is chordal, it admits a representation as a tree of cliques, which can be converted into a tree of complexes in the original graph by deleting the artificially-added edges. This method was applied to the TNF-$\alpha$/NF-$\kappa$B and pheromone signaling pathways~\cite{zotenko2006decomposition}. To further the use of hypergraphs in network biology, it will be important to generalize this method to apply to larger classes of graphs and to unify these methods of automated reconstruction with the results of manual curation. It may also be valuable to formulate hybrid network representations that combine the features of pairwise graphs and hypergraphs. A caveat here is that the need to develop a novel set of algorithms for every new representation might prevent its wide adoption in the community.

The second issue is that the theory for (directed) hypergraphs is much less well-developed than for pairwise graphs. Problems that have well-established and simple polynomial-time solutions on pairwise graphs, e.g., shortest paths, turn out to be computationally intractable on directed hypergraphs~\cite{ritz2014signaling}, as discussed above. Incorporating regulation into the definitions of shortest paths continues to be challenging~\cite{krieger2022computing}. Moreover, graph-theoretic concepts such as clusters, flows, random walks, or convolutions that have been employed fruitfully in network biology are either challenging to generalize to hypergraphs or have found limited applications in biology.  

The third issue is that it is not clear under what conditions or for which applications a higher-order representation is better than a pairwise graph representation. Arguments often appeal to visual and qualitative reasoning (Fig.~\ref{fig:TGFb}). We encourage the community to come forward with well-established datasets, evaluation measures, and benchmark frameworks that can pose these questions formally and develop generalizable standards. 

%% file: 050.machine_learning_rtf.tex
\xhdr{Overview}
Machine learning has emerged as a powerful paradigm for creating predictive models specified as parameterized functions with tunable parameters that operate on structured data, such as graphs, spatial geometries, relational structures, and manifolds. Applying machine learning methods to network data has demonstrated potential in a myriad of biological network analysis tasks~\cite{hetzel2021graph,li2022graph,theodoris2023transfer,yue2020graph}. Recent methods are designed to produce graph representations as compact numerical vectors (or embeddings) corresponding to various graph elements, such as nodes, edges, subgraphs, and entire graphs, and capture essential information about the topology of these elements. These learned representations can be fed into models trained toward a vast array of downstream analytic tasks. 

\begin{figure*}[t!]
  \centering
   \includegraphics[width=\linewidth]{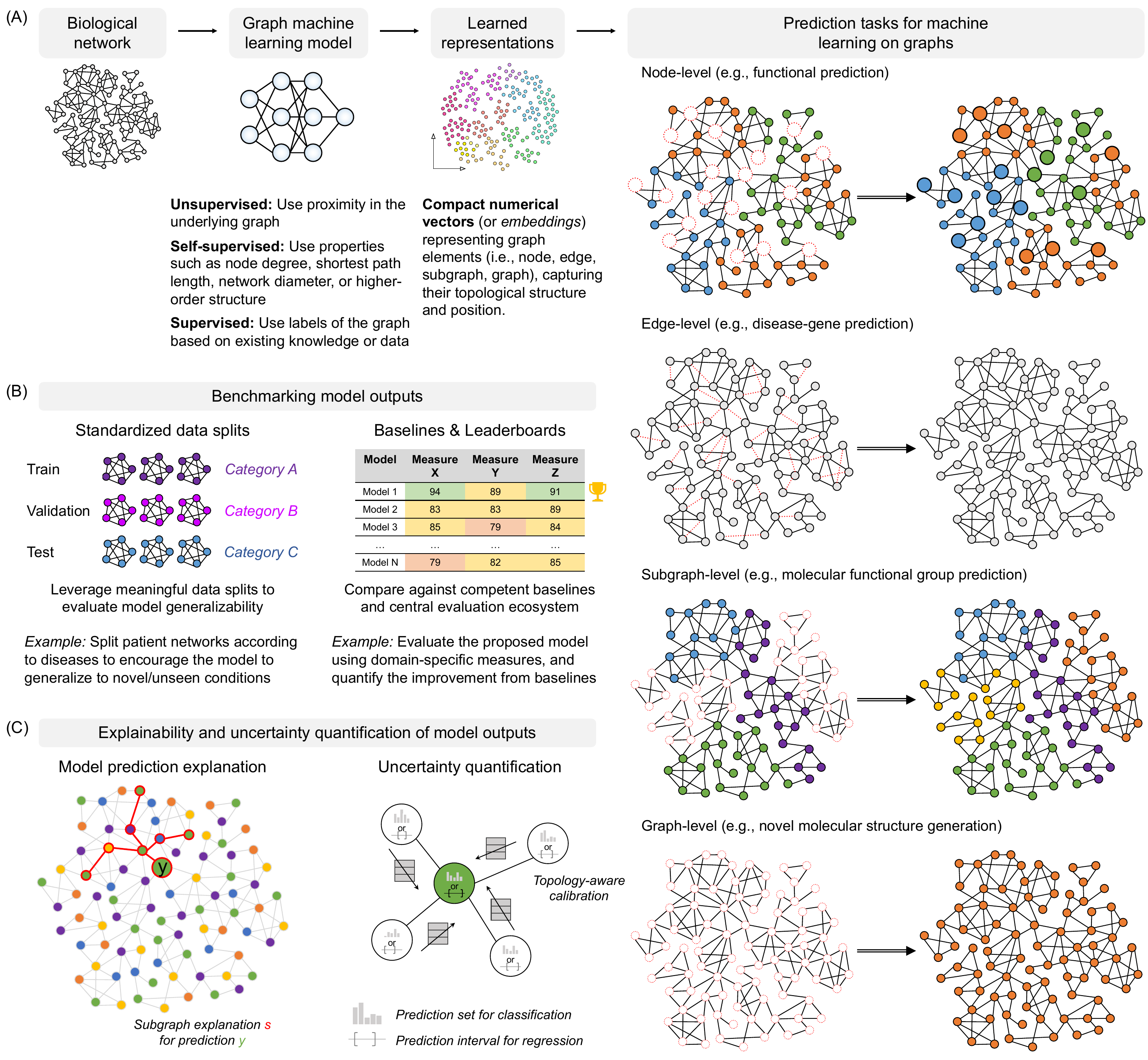}
  \caption{{Overview of the components of machine learning on networks. \textbf{(A)} The core of this approach is a machine learning model, typically a neural network, that takes one or more biological networks as input and learns representations (i.e.,~embeddings) of various graph elements in an unsupervised, self-supervised, or supervised manner. There are four types of prediction tasks (denoted by the red dashed lines): node-, edge-, subgraph-, and graph-level predictions. Colors of nodes for the node-, subgraph-, and graph-level tasks signify the label; white nodes indicate missing labels to be predicted by the model. Examples include functional prediction (node-level), disease-gene prediction or context-specific edge prediction (edge-level), molecular functional group prediction (subgraph-level), and novel molecular structure generation (graph-level). Critical to continued development, wide adoption, and practical utility of network-based machine learning is a parallel improvement in frameworks for \textbf{(B)}~rigorous benchmarking via established data splits and baselines, and \textbf{(C)}~explainability of model predictions (e.g., identifying a subgraph $s$, denoted by red lines, that best explains the prediction $y$ for the query node, denoted in green) and uncertainty quantification (e.g., using the prediction set for a classification task or prediction interval for a regression task~\cite{huang2023uncertainty}).}}
  \label{fig:ML_networks}
\end{figure*}

{Predictive models on graphs include models for predicting node labels (node classification), edge-level relationships (link prediction), subgraph-level labels (subgraph classification), and graph-level labels (graph classification) (Fig.~\ref{fig:ML_networks}). These models} can be created through unsupervised, self-supervised, and supervised learning on all types of networks, including homogeneous, heterogeneous, temporal, and spatial networks, and with additional constraints and domain knowledge imposed on the models. By leveraging deep graph learning models pretrained on large-scale general graph datasets, it is possible to adapt (or fine-tune) pretrained representations for diverse use cases in predictive and generative modeling~\cite{gainza2020deciphering,gainza2023novo}. As machine learning on graphs continues to be developed, appropriate model benchmarking is necessary to ensure that task-specific evaluation measures are well-defined and predictions are fair and robust. The rest of this section discusses these topics, which are also summarized in Fig. \ref{fig:ML_networks}.

\begin{table*}[ht]
\caption{{Prominent open-source benchmark datasets for machine learning on biological networks. Databases are categorized by data type. The table is organized alphabetically by data type and database names.}}
\label{tab:ml-resources}
\begin{center}
\begin{small}
\begin{tabular}{p{0.09\linewidth}p{0.3\linewidth}p{0.09\linewidth}p{0.43\linewidth}}
\toprule
\textbf{Data type} & \textbf{Database} & \textbf{Task type} & \textbf{Prediction tasks}  \\

\midrule
General & Long Range Graph Benchmark~\cite{dwivedi2022LRGB} & Edge-level & Molecular bond \\
& &  Graph-level & Peptide function, peptide structure \\

\midrule
General & Open Biomedical Network & Node-level & Protein function \\
& Benchmark~\cite{liu2023nleval} & Edge-level & Disease-gene association \\

\midrule
General & Open Graph Benchmark~\cite{hu2020open} & Node-level & Protein function \\
& & Edge-level & Protein-protein association, drug-drug interaction, heterogeneous interaction, vessels in mouse brain \\
& & Graph-level & Molecular property, species-specific protein association \\

\midrule
General & SubGNN Benchmarks~\cite{alsentzer2020subgraph} & Subgraph-level & Proteins associated with biological process, rare neurological disorders phenotype-based diagnosis, and rare metabolic disorders phenotype-based diagnosis \\

\midrule
General & Temporal Graph Benchmark~\cite{huang2024temporal} & Node-level & Dynamic node affinity prediction \\
& & Edge-level & Dynamic link prediction \\

\midrule
Knowledge graph & PrimeKG~\cite{chandak2023building} & Node-level & Identity of protein/gene, disease, drug, biological process, pathway, phenotype, molecular function, cellular component, exposure, and anatomical region \\
& & Edge-level & Protein-protein interaction, disease-drug indication, disease-drug contraindication, disease-drug off-label use, disease-phenotype association, disease-disease association, disease-protein association, disease-exposure association, phenotype-protein association, pathway-gene association, etc. \\

\midrule
Knowledge graph & Phenotype Knowledge Translator~\cite{callahan2024open} & Node-level & Identity of tissue, cell, DNA, RNA, gene, miRNA, variant, protein, disease, biological process, pathway, phenotype, molecular function, cellular component, and chemical \\
& & Edge-level & Tissue-/cell-specific gene expression, gene-variant association, variant-disease association, chemical-disease association, chemical-pathway association, etc \\ 

\midrule
Molecular & Protein sEquence undERstanding~\cite{xu2022peer} & Edge-level & Protein-protein interaction, contact prediction \\
design & & Graph-level & Molecular property (e.g., fold classification, secondary structure prediction) \\

\midrule
Molecular & Tasks Assessing Protein Embeddings~\cite{tape2019} & Edge-level & Protein-protein interaction, contact prediction \\
design & & Graph-level & Molecular property (e.g., fold classification, secondary structure prediction) \\

\midrule
Molecular design & Graph Explainability Library~\cite{agarwal2023evaluating} & Graph-level & Molecular mutagenic property, molecular functional group (e.g., benzine rings, fluoride carbonyl) \\

\midrule
Neurology & NeuroGraph~\cite{said2023neurograph} & Graph-level & Donor demographics (age and gender), task states (emotion processing, gambling, language, motor, relational processing, social cognition, and working memory), cognitive traits (working memory, fluid intelligence) \\

\midrule
Therapeutic discovery & AVIDa-hIL6~\cite{tsuruta2024avida} & Edge-level & Antigen-antibody interaction \\

\midrule
Therapeutic discovery & Therapeutic Data Commons~\cite{huang2021therapeutics} & Edge-level & Drug-target interaction, drug-drug interaction, protein-protein interaction, disease-gene association, drug-response prediction, drug-synergy prediction, peptide-MHC binding, antibody-antigen affinity, miRNA-target prediction, catalyst prediction, TCR-epitope binding, and clinical trial outcomes \\
& & Graph-level & Molecular property (e.g., synthesizability, drug-likeness) \\

\bottomrule
\end{tabular}
\end{small}
\end{center}
\end{table*}

\xhdr{Unsupervised, self-supervised, and supervised graph learning}
Unsupervised learning of graph representations involves optimizing parameterized strategies, such as GNNs, graph transformers, or multi-layer neural message-passing models, to aggregate information from a node's (e.g., a gene in a gene co-expression network or a patient in a patient similarity network) neighbors in the network. The goal is to optimize the representations so that the proximity between entities in the embedding space mirrors their proximity in the network~\cite{atz2021geometric,cao2020comprehensive}. Prevalent strategies for sampling neighbors in the network vicinity of nodes that get embedded in the latent space include biased and unbiased random walks as well as adaptive neighbor sampling~\cite{hamilton2017inductive,velivckovic2018deep}. Objective functions of these methods aim to maximize embedding similarity in the latent space for neighboring nodes in the network~\cite{hamilton2020graph,hamilton2017representation,perozzi2014deepwalk,tang2015line}. For instance, nodes connected by edges should be embedded closer together in the latent space (i.e., have more similar embeddings) than nodes that are not connected~\cite{grover2016node2vec,liu2022graph,wu2021self,xie2022self}. 

{Self-supervised graph representation learning, the predominant approach for machine learning on graphs, leverages not only the network structure but also additional context or auxiliary tasks to generate informative embeddings. Unlike unsupervised methods that solely rely on the network structure for optimization, self-supervised techniques utilize auxiliary (pretext) tasks, such as predicting node attributes or reconstructing graph substructures, to enhance the learning process and create more robust embeddings~\cite{hassani2020contrastive,li2022graph,zitnik2018modeling,zitnik2017predicting}.} An example of a self-supervised node-level auxiliary task is predicting each node's degree. Link prediction is a self-supervised edge-level task that predicts whether an edge exists between a pair of nodes~\cite{kipf2016variational,li2022graph} based on a self-supervised objective~\cite{liu2021self}, which can be formulated using contrastive learning~\cite{you2020graph}, node or edge masking~\cite{agarwal2023evaluating}, and generative denoising~\cite{yi2024graph}. Examples of self-supervised subgraph and graph tasks include predicting subgraph and graph properties, such as distributional statistics of shortest path lengths, network diameter, and the presence or absence of specific higher-order structures and graphlets~\cite{alsentzer2020subgraph,luo2022clear,you2020graph}.

Graph representation learning, whether unsupervised or self-supervised, can be applied to any type of network, including but not limited to homogeneous, heterogeneous, temporal, spatial, and physical networks. For example, in heterogeneous networks, GNN and graph transformer models leverage node- and edge-based attention weights to aggregate neighborhood information depending on node and edge types~\cite{fu2022mvgcn,kesimoglu2023graf, wang2019heterogeneous, xie_mgat_2020,Zhang2019HetGNN}. Other approaches treat each edge type as a homogeneous graph, apply a graph representation learning model to it, and then integrate edge-type specific node representations into final representations~\cite{fu2022mvgcn, kesimoglu2023graf,kesimoglu2022supreme, wang2021mogonet}. {In a heterogeneous network, subgraphs can be sampled via metapaths~\cite{sun2011pathsim}, which are defined by sequences of relationships (or edge types) connecting different types of nodes to model semantic nuances underlying the network in a self-supervised manner, such as through contrastive learning~\cite{dong2017metapath2vec,zhao2021multi}.} These advancements in graph representation learning have impacted areas like cancer biology, drug discovery, and disease diagnosis~\cite{esteva2019guide,huang2023zero,huang2022artificial,morselli2021network,stokes2020deep}.

{Supervised graph representation learning uses networks with additional expert-curated or experimentally-derived labeled data to directly optimize models for specific prediction tasks (Fig.~\ref{fig:ML_networks}A).} In this paradigm, nodes, edges, subgraphs, or entire graphs are associated with labels, and the learning process minimizes the discrepancy between the model's predictions and these labels~\cite{Schlichtkrull2018,velivckovic2017graph}. Common applications include node classification, where individual nodes are assigned to predefined categories, and graph classification, wherein entire graphs are categorized based on their topological  features~\cite{eyuboglu2022mutual,gilmer2017neural}. Unlike unsupervised and self-supervised models, supervised graph learning directly uses label information, often leading to more task-specific and accurate representations, albeit at the cost of requiring labeled data.

\xhdr{Incorporating knowledge into machine learning models through knowledge graphs, spatial constraints, equivariances, and symmetries} 
In numerous biological and medical applications, standard graph representation learning often falls short of requirements. In these cases, the model's predictive accuracy can be enhanced by imposing constraints drawn from pre-existing knowledge. Typical strategies encompass incorporating multimodal data into BKGs, augmenting GNNs with bespoke architectures, and applying domain-specific invariances.

BKGs help model heterogeneous relationships between biomedical entities, as already discussed in Section \ref{sect:multimodal_networks_integration}. The resulting latent space, which reflects the topology of the underlying knowledge graph, can be operated on to make inferences about existing and novel relationships. Jointly modeling diverse types of relationships in a BKG, such as integrative modeling of transcription regulation and metabolism~\cite{chandrasekaran2010probabilistic,niu2021trimer}, can present unique challenges due to the BKG's incompleteness and potential high-order relationships involving heterogeneous entities. Incorporating pathway knowledge, either implicitly as constraints that regularize network embeddings~\cite{niu2021trimer} or directly as a prior placed on the BKG structure and parameters in a Bayesian fashion~\cite{boluki2017incorporating}, has been shown to improve predictive performance. Supervised machine learning methods often require many samples to identify biologically meaningful patterns, which can limit their applicability in areas such as rare diseases that are inherently limited in clinical cases, leading to few samples to analyze~\cite{banerjee2023machine}. Advances in self-supervised graph learning applied to BKGs have shown promise for rare disease research~\cite{alsentzer2022deep} and will likely be informative for applications beyond rare diseases for which few samples exist with high-dimensional data.

Temporal and spatial data can be represented as networks, but specialized neural architectures are necessary to learn optimally on temporal/dynamic networks. Temporal graph representation learning methods typically involve two main components: a GNN architecture to generate embeddings for each time point and a recurrent neural network, such as a long short-term memory network or a transformer network, to perform sequence learning by leveraging temporal relationships between elements in the sequence. Existing approaches use GNNs as feature extractors of nodes and the underlying topology, and recurrent neural networks for temporal learning and to include additional metadata information~\cite{li2017diffusion, manessi2020dynamic,pareja2020evolvegcn,peng2020spatial, zhao2019t}. Recently, static GNNs have been extended to handle dynamic graphs by treating time points as hierarchical states~\cite{you2022roland} or applied to irregular time series data by propagating neural messages between time intervals of each sensor as well as between sensors~\cite{zhang2022raindrop}. Protein molecular configurations can be depicted as protein structure networks where amino acid nodes are linked by the 3D physical proximity of their residues, and the amino acid spatial coordinate information is encoded as node attributes. Deep learning models, particularly through the use of equivariant GNNs, can both attain high performance and preserve transformations of protein networks under translation, reflection, and rotation of networks in the 3D space~\cite{batzner20223,gong2023general,jumper2021highly}. For instance, to establish a model that remains invariant to molecular spatial orientation, constraints enforcing rotation invariance ought to be integrated~\cite{jumper2021highly}. Methodologies derived from equivariant neural networks, such as AlphaFold~\cite{jumper2021highly}, can complement sequence-based language models~\cite{lin2023evolutionary} by harnessing evolutionary data to infer protein structures from primary amino acid sequences, and potentially generate realistic molecular formations.

\xhdr{Generative graph models}
Generative graph models are a class of machine learning models specifically designed to generate new graphs, or parts of graphs, that resemble {a given set of training graphs in some way. These models learn to capture the underlying patterns and structures in the training graphs} and can then be used to produce new graphs with similar properties as the training graphs. For example, in molecular biology, the inherently graph-like nature of molecular structures has made GNNs an ideal tool for generating drug-like molecules, guiding the generation process by learning the underlying patterns and properties from real molecular data~\cite{bilodeau2022generative}. One such method is a variational graph autoencoder that learns embeddings of molecular structures and uses them to generate novel molecular graphs~\cite{jin2018junction,kipf2016variational, li2018learning}. Other generative models, such as GraphVAE, GraphRNN, and MolGAN, have also been developed to generate realistic graphs~\cite{de2018molgan,simonovsky2018graphvae, you2018graphrnn}. Inspired by generative adversarial networks for image generation, MolGAN pits a generator model (which produces graphs) against a discriminator model (which tries to distinguish between real and generated graphs). Additionally, graph transformer networks have recently been proposed for molecular graph generation, demonstrating the ability to generate molecules with desired properties by training on extensive chemical databases~\cite{bagal2021molgpt}.

When applied to protein design, GNNs have demonstrated impressive results in designing protein sequences that fold into specific structures~\cite{ingraham2019generative}. Graph-based methods like PotentialNet have shown promise for protein-ligand binding prediction~\cite{feinberg2018potentialnet}. Similarly, DeepSite uses 3D convolutional neural networks to predict protein-ligand binding sites~\cite{jimenez2017deepsite}. Moreover, recent generative models, such as ProteinMPNN~\cite{dauparas2022robust} utilize message-passing neural network architecture to generate protein sequences and structures, further expanding the range of possibilities for protein design.

{Diffusion models have recently emerged as powerful tools in protein and drug design~\cite{abramson2024accurate, corso2023diffdock,watson2023novo,yim2024diffusion}, leveraging their capability to model complex distributions for generating novel molecular and protein structures. In protein design, diffusion models operate by gradually denoising a random configuration towards a target protein structure, learning the distribution of protein conformations. A notable example is RFDiffusion~\cite{watson2023novo}, a diffusion model that generates protein structures by conditioning on both sequence and structural information, achieving enhanced accuracy in structure prediction. In drug design, these models are adapted to generate molecular graphs by iteratively refining a random molecular graph into a drug-like molecule with desired properties through a learned diffusion process~\cite{o20243d}. }

\xhdr{Transfer learning}
The quality of representations generated by graph representation learning methods is contingent upon the availability of labels. Nevertheless, in the realm of network biology, labels are often in short supply due to the substantial resources required for their curation and validation. A potent solution to addressing this challenge is transfer learning. This approach involves initially training a graph representation learning model on a large reference network via self-supervised pretraining~\cite{hu2020strategies,li2022graph,xie2022self, you2020graph}, followed by adapting the resulting model or its outputs to a different task of interest typically through supervised learning on a small set of labeled examples (fine-tuning). Pretraining a model on a large network followed by fine-tuning of the model using a small labeled dataset allows the model to harness extant information about a network entity (i.e., from the large network utilized for pretraining) in service of diverse tasks with limited task-specific labels.

Transfer learning has shown considerable potential for developing predictive models on condition-specific networks that vary with biological conditions. Networks are typically constructed from context-{unaware} data (e.g., the human reference PPI network~\cite{luck2020reference}) or data generated under specific conditions (e.g., a gene co-expression network for a particular disease). Biomedical entities and their interactions can vary across biological conditions, such as tissues, cell types, and disease states. Nevertheless, generalizing knowledge from context-{unaware} networks to context-specific problems presents considerable challenges. {For instance, modeling tissue- or cell type-specific interactions from the human reference PPI network requires the construction of tissue- and cell type-specific networks and the development of multi-scale network models~\cite{greene2015understanding, ietswaart2021genewalk, li2021deep,zitnik2017predicting}.} One approach to this challenge involves constructing context-specific networks (as discussed in Sections \ref{sect:network_inference_comparison} and \ref{sect:multimodal_networks_integration}) and applying independent shallow network embedding layers to learn node representations based on network topology and tissue hierarchical structure~\cite{greene2015understanding, zitnik2017predicting}. An alternative strategy is to learn shallow network embeddings on a context-{unaware} network, such that the embeddings of nodes operating in the same context are more similar to each other than nodes operating in different contexts~\cite{ietswaart2021genewalk}. Recent methods incorporate context in a data-driven manner, constructing cell type-specific PPI networks using single-cell transcriptomic data~\cite{li2023contextualizing,li2021deep}. Unified by a network of cell type and tissue hierarchy, these networks can be harnessed to learn unique protein representations tailored to each cell type context~\cite{li2023contextualizing,li2021deep}.

\xhdr{Understanding predictive models, benchmarking, and rigorous evaluation across diverse tasks}
With the rapid evolution of graph learning methodologies, the need to construct rigorous benchmarks for effectively assessing the performance of these novel techniques is becoming increasingly urgent~\cite{hu2020open,shchur2018pitfalls}. Open-science evaluation platforms such as the Benchmarking GNN~\cite{dwivedi2020benchmarking}, Open Graph Benchmark~\cite{hu2021ogb,hu2020open}, {and others (Table \ref{tab:ml-resources})} serve as significant assets for general graph benchmarking, while other resources are being curated explicitly for the domain of network biology~\cite{liu2023nleval}.

To provide a comprehensive evaluation, these resources ought to be expanded to include tasks defined at various levels of graphs, including node classification, link prediction, subgraph classification and clustering, and whole-graph classification and regression. In addition to benchmarking models for predictive tasks, evaluation frameworks are needed for generative graph models. They should also encompass diverse types of biological graphs, such as heterogeneous, spatial, and temporal ones. {A critical element in this regard is benchmarking the performance of network-based machine learning techniques across multiple dimensions of evaluation beyond accuracy, including robustness, generalizability, and computational efficiency.} 

Moreover, the explainability of graph-based learning can offer significant insights in the biomedical domain~\cite{agarwal2023evaluating,xie2022task,ying2019gnnexplainer,yuan2021explainability}. Consequently, it is equally important to examine learned algorithms by examining pretrained graph representations~\cite{forster2022bionic} and mapping attention mechanisms in attention-based deep learning models~\cite{elmarakeby2021biologically}. As we move towards the broader application of machine learning models in network biology, proper quantification of the uncertainty, error, and utility associated with these models is indispensable. Given the potential for considerable uncertainty in these models, effective techniques for uncertainty quantification are required to fully comprehend the predictive capabilities and limitations of a given model~\cite{abdar2021review}.

When the model's objective is specific, such as treatment recommendation, disease diagnosis and prognosis, and steady-state or transient network behavior prediction, an objective-driven approach to uncertainty quantification can be beneficial~\cite{yoon2013quantifying}. This approach allows us to quantify uncertainty based on its impact on the expected performance of prediction and intervention tasks. Ultimately, this can pave the way for optimal experimental design techniques~\cite{dehghannasiri2014optimal,dehghannasiri2015efficient} that prioritize experiments to generate the most informative data points selected by active learning strategies, effectively reducing model uncertainty.

%% file: 060.personalized_medicine_rtf.tex
\xhdr{Overview} The overarching goal of precision medicine is to develop diagnostic and treatment strategies tailored to individual patients \cite{aronson2015building,kaiser2015nih,NMD18}, while also taking into account the desired level of precision for each treatment. Personalized characterization of an individual or a group can encompass various data types, including molecular, healthcare, environmental, lifestyle, and behavioral information, commonly modeled and analyzed as networks \cite{PrzuljSci2016}. By assimilating data from different modalities, precision therapeutics can amplify their potential and bolster resilience against diverse data noise~\cite{Glig2016,huang2021therapeutics,huang2022artificial}. Fusing data from multiple sources has proven effective in advancing precision medicine \cite{Thomas21,PSB2016,iCell2019,wang2014similarity}.

\begin{figure*}[t!]
  \centering
    \includegraphics[width=1\linewidth, trim=0cm 0.4cm 0cm 0cm]{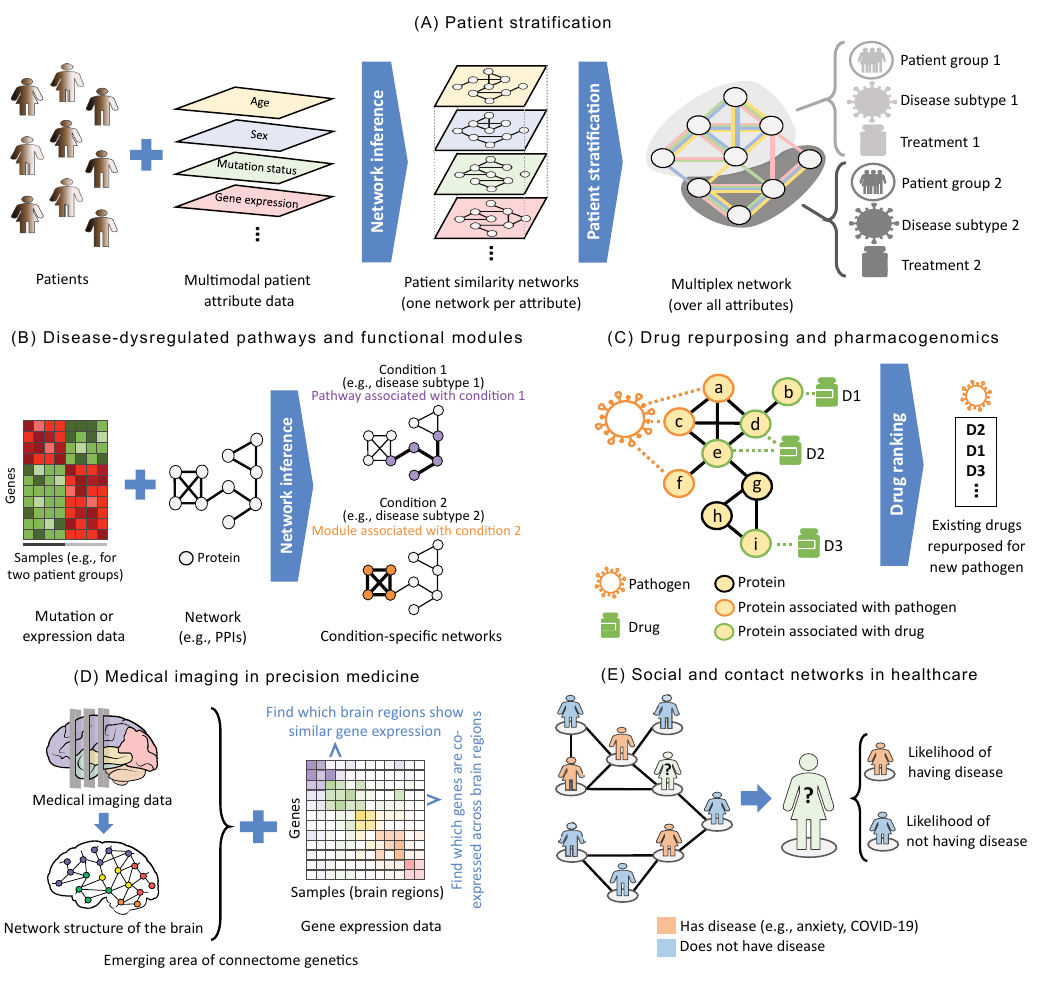}
    \caption{Prominent topics in network-based precision medicine. \textbf{(A)} Groups of patients that correspond to their communities (clusters) in a patient similarity network may shed light on distinct disease subtypes and thus lead to tailored, group-specific therapeutic strategies. \textbf{(B)} Identification of pathways (sparse, tree-like subnetworks) or functional modules (dense, clique-like subnetworks) associated with disease (subtypes) is related to inference of a condition-specific network (Section \ref{sect:network_inference_comparison}) and pathway reconstruction (Section \ref{sect:multimodal_networks_integration}). \textbf{(C)} Drug repurposing evaluates the fit of existing drugs to new diseases based on network ``relatedness'' between protein targets of the existing drugs and proteins associated with the new diseases. E.g., existing drug D2 may be a good treatment for the new pathogen because D2 targets two proteins (d and e), both of which directly interact with two of the proteins associated with the pathogen (a and c); the four proteins (a, c, d, e) form a clique, which further adds to their ``relatedness''.  \textbf{(D)} An important application of medical imaging lies in brain disorders. In connectome genetics, network structure of the brain meets -omics data.  \textbf{(E)} An individual's position in their social/contact network, along with demographic, personality, physical/mental health, etc. information about the other individuals, can give insights into the given individual's health.}
   \label{fig:precision_medicine}
\end{figure*}

\xhdr{Patient stratification}
Precision medicine aims to provide individualized diagnostic and treatment strategies. Developing treatments tailored to specific patient groups based on distinct disease subtypes (Fig. \ref{fig:precision_medicine}A) is poised to transform a prevailing one-size-fits-all approach used in healthcare. Network methods can integrate multimodal data to identify patient groups with coherent genetic, genomic, physiological, and clinical profiles~\cite{ektefaie2023multimodal,PSB2016,petti2023network}, even when the underlying data are incomplete and noisy \cite{pai2018patient}. The methods assume that patients with similar clinical signatures and similar -omics profiles have similar clinical outcomes. Similarities between patients can be efficiently represented through patient similarity networks; in these networks, nodes symbolize patients, and weighted edges denote the degree of similarity derived from clinical and biomolecular patient attributes. Each patient data attribute, such as age, sex, mutation status, or gene expression profile, can be used to create a network of pairwise patient similarities. Then, the set of all such networks can be viewed as a multiplex network, with a  layer for each of the attributes. Various similarity measures can be employed to assess patient similarity across different datasets corresponding to different attributes. After the multiplex patient similarity network is constructed, patient subtypes can be identified by examining the community (clustering) structure within the network.  Communities are characterized as subsets of nodes that are densely connected to each other and loosely connected to nodes in different communities {\cite{Fortunato2010}}. Communities in a patient similarity network are thus densely/strongly linked patient groups and can shed light on distinct disease subtypes.

Network methods offer distinct advantages over non-network approaches, which often grapple with the complexities of integrated datasets \cite{gligorijevic2015methods}. Patient stratification has increasingly benefited from network-based methodologies, which can elucidate intricate biological interactions, especially within disease mutation landscapes, such as cancer \cite{PSB2016} or rare hereditary diseases \cite{Thromb23}. {By studying different types of gene-gene interactions, encompassing aspects like mutual exclusivity, co-occurrence, and both physical and functional associations, and analyzing personalized gene regulatory networks~\cite{rogers2022network}, one can better understand inter-individual variation in disease driven by differences in interactions caused by each patient’s genetic background, environmental exposures, and the proportions of specific cell types involved in disease~\cite{van2018integrative}.} Such insights can elevate the accuracy of patient stratification, which is typically measured as the ability to classify patients as belonging to known disease subtypes~\cite{pai2019netdx} or the ability to identify disease biomarkers that generalize (maintain performance) when applied to new data that have not yet been seen by the model~\cite{alsentzer2022deep,kong2022network}. These insights can also guide the refinement of therapeutic strategies, ensuring they are optimally tailored to specific patient groups~\cite{dao2017bewith,PSB2016,huang2023zero}. 

\xhdr{Identification of pathways associated with disease subtypes and patient groups} 
Identifying group-specific mutations provides valuable insights into the underlying biochemical pathways associated with the disease (Fig. \ref{fig:precision_medicine}B). These pathways can be conceptualized as networks, laying the foundation for an in-depth understanding of disease mechanisms. Incorporating individual mutation or expression data into pathway-based (i.e., network-based) methods aid in identifying targetable mutations \cite{park2019pathway}. This approach is especially pertinent in determining functional pathways that play roles in expression responses to disease-propagating mutations, leveraging the concept of pathway centrality \cite{Sam22A,Sam22}.

For instance, by integrating genomic, clinical, and therapeutic data through networks, physicians can categorize patients with treatment-resistant prostate cancer based on specific gene mutations like AR, PTEN, and BRCA2. Recognizing these mutations facilitates the adoption of personalized therapies, targeting the aberrant pathways distinctive to each patient's tumor profile. As a result, this tailored treatment strategy offers the potential for safer and more effective treatments \cite{mateo2020accelerating}.

Furthermore, recent research has illuminated the importance of tissue-specific regulatory networks and the pathways they encompass, which frequently manifest genetic mutations in particular patient cohorts. This understanding emerged from the combined analysis of expression and chromatin accessibility data, unveiling a previously unidentified tissue-specific stem-cell-like subtype of treatment-resistant prostate cancer that may be a target for intervention \cite{tang2022chromatin}. Similarly, a comparative structural analysis of the chromatin structure network in chronic lymphocytic leukemia and control tissue of origin revealed that genes driving this cancer type are characterized by specific local wiring patterns not only in the chromatin structure network of chronic lymphocytic leukemia cells but also of healthy cells \cite{CLL20}. This allows for the successful prediction of new DNA elements related to this cancer type, and importantly, it shows that cancer-related DNA elements can be identified in other cancer types by investigating the chromatin structure network of the healthy cell of origin, a critical new insight paving the road to new therapeutic strategies \cite{CLL20}.

\xhdr{Identification of disease-dysregulated functional modules}
Studying disease-dysregulated functional modules of genes can advance the understanding of disease beyond isolated mutations or pathway dysregulations. Disease-associated behaviors can materialize in clusters of tightly interacting proteins forming functional modules (Fig. \ref{fig:precision_medicine}B) ~\cite{agrawal2018large,menche2015uncovering} rather than exclusively via singular gene mutations or perturbed gene expression \cite{schadt2009molecular}.

The quest to uncover disease-associated functional gene modules from molecular networks is a long-standing challenge with implications for precision medicine \cite{barabasi2011network,choobdar2019assessment,eyuboglu2022mutual,Thomas20,mitra2013integrative, gysi2022non}. Prevailing approaches for finding disease modules rely on the assumption that interacting genes tend to associate with similar phenotypes. For instance, gene co-expression network analysis has been employed to pinpoint modules of genes that exhibit analogous co-expression patterns in breast cancer. Notably, these clusters of genes correlate with distinct metastasis progression patterns in patients \cite{chuang2007network}. {Multi-omic module detection in cancer can consider mutation mutual exclusivity, transcriptional regulation, and gene co-expression alongside PPI connections \cite{silverbush2019simultaneous}}.

{Given the complexity of disease circuits in many complex diseases, concentrated efforts have been directed toward identifying disease-associated gene modules that correlate with patient phenotypes~\cite{choobdar2019assessment,saelens2018comprehensive}. Disease-associated gene modules, identified through computational approaches and various types of gene networks, have been used to refine disease diagnosis \cite{morselli2020whole}.} They can also forecast the response of individual cell lines to specific anticancer agents and potentially suggest patient-tailored drug combinations \cite{kim2020identifying,Salazar21}. Supplementing these techniques, differential network analysis (Section \ref{sect:network_inference_comparison}) can reveal differential connections or rewiring of a molecular network under varying conditions. This complements traditional differential gene expression analyses, giving a robust framework to investigate diverse conditions and, by extension, different patient groups \cite{gysi2020construction,morselli2020whole,tu2021differential}.

Precision medicine's applications in identifying candidate anticancer therapeutics have broadened its scope to probe molecular shifts linked with other diseases and aging. Recent research endeavors have used multi-omics strategies to pinpoint innovative therapeutic targets for ulcerative colitis \cite{voitalov2022module} and rheumatoid arthritis \cite{li2023contextualizing}. As another example, complementing the above discussion of detecting disease-associated modules of genes from a molecular network, modules of diseases have been detected from a heterogeneous disease-disease similarity network \cite{halu2019multiplex}. Other studies have delved into molecular biomarkers, their regulatory pathways, and age-related modifications \cite{tseng2018peripheral}. These studies aim to formulate therapies adeptly tailored to diverse age demographics. Complementing the focus on aging, there is a burgeoning interest in discerning patient sex-specific disparities. These lines of inquiry draw motivation from epidemiological data, which delineate differential patterns in the incidence, progression, and prognosis of complex diseases across gender and age brackets \cite{cannistraci2021age}.

\xhdr{Drug repurposing and pharmacogenomics}
Compared to traditional drug development, drug repurposing (Fig. \ref{fig:precision_medicine}C) offers significant advantages such as low cost, reduced risk, and faster drug development {timelines~\cite{cheng2018network,langhauser2018diseasome,pushpakom2019drug,unsal2023nmsdr}}. While early examples of successfully repurposed drugs have been identified through serendipitous discoveries, the availability of massive amounts of -omics and knowledge data and advances in computational techniques have provided opportunities for systematic \emph{in silico} inference of novel indications for existing drugs~\cite{guney2016network,huang2023zero,wen2023multimodal,Xenos23,Carme21}. Network science and machine learning models have demonstrated impressive capabilities, but the bar for clinical applications is high. For example, an ensemble network approach has been used to identify drug candidates for repurposing against COVID-19 viral replication \cite{morselli2021network,patten2022identification}. As another example, a heterogeneous network approach revealed diseases that are most similar to COVID-19, thus reflecting conditions that are risk factors in patients and suggesting the suitability of this approach for use in drug repurposing \cite{Verstraete2020CovMulNet19}. Validation of the most promising computational predictions in the laboratory yielded an order of magnitude more potent candidates than non-guided experimental screening. In pharmacogenomics, graph convolutional neural networks trained on heterogeneous networks of drug-drug interactions identified adverse events due to polypharmacy and concomitant use of medications \cite{zitnik2018modeling}. Furthermore, deciphering drug-cell connectivity data, indispensable for patient-specific drug repositioning, gains momentum by embedding PPI networks using tensor completion algorithms \cite{bumin2022fit}.

\xhdr{The role of medical imaging in precision medicine}
In addition to -omics data, medical images have emerged as an important new data modality that can facilitate precision medicine, including disease detection, diagnosis, and therapeutic interventions \cite{comaniciu2016shaping,lambin2017radiomics}. Often, medical images encompass distinct topological patterns of target entities that can serve as diagnostic signatures or biomarkers, such as the dendritic structure of the trachea or clustering behaviors of immune cells. Combining these topological signatures with deep learning algorithms offers a substantial advantage in various medical image analysis endeavors, including segmentation, classification, registration, and tracking, and can help with the interpretability of deep learning models. Building tools to compute topological and deep learning representations of imaging data inaugurates new avenues for nuanced analysis, unveiling hidden patterns and intricate correlations within multifaceted datasets \cite{edelsbrunner2002topological}. These developments have catalyzed the birth of topology-infused deep learning techniques for myriad applications, spanning from segmenting retinal vessels \cite{hu2019topology,shit2021cldice} to discerning retinal arteries/veins \cite{mishra2021vtg} and forecasting protein semantic similarities \cite{wang2022tango}.

An important application of network-based precision medicine lies in brain disorders, where medical image analysis intertwines with network and -omics data (Fig. \ref{fig:precision_medicine}D). Specifically, {procuring multimodal neuroimaging, neural network configurations, genetic markers, and other biomolecular signatures could allow for gaining insights} into the neural architectures of the human brain, the modulation of its functionalities by network topographies, and the genetic interplays that correspond to disease-specific cerebral patterns. An emergent discipline, dubbed connectome genetics, heralds the meticulous delineation of human neural connectivity, unraveling its ties to cognition, behavior, and the genetic underpinnings of individual neural circuit variances \cite{arnatkeviciute2021genome}. Graph mining techniques combined with data science methods have been devised, geared towards personalizing diagnosis and therapy by leveraging the multifaceted data from connectome genetics \cite{arnatkeviciute2021genetic,jahanshad2013genome,sha2023genetic}. The recent advent of GNN-driven deep learning models further deepens our grasp on the intricate shifts within this data, advancing our understanding of neurological diseases and their heterogeneity across patient populations \cite{zhang2019brain,zhang2021disentangled,zhao2022revealing}.

\xhdr{The role of social and contact networks in healthcare}
Biological networks hold significant promise for advancing personalized medicine. In tandem, social, support, and contact networks correlate with individual health outcomes (Fig. \ref{fig:precision_medicine}E), providing valuable insights into patient behaviors and sentiments \cite{smith2008social}. Such networks offer real-time perspectives on patient inclinations, such as therapy adherence preferences. Moreover, they can model patient behaviors associated with medication consumption, enabling the formulation of individualized intervention strategies \cite{guinazu2020employing}. The confluence of health and social networks has been harnessed to forecast individual health outcomes, including mental health parameters like anxiety and depression. These predictions emerge from a rich tapestry of data sources, including combinations of heterogeneous social network data and wearable health measures \cite{liu2021heterogeneous}, and dynamic social network interactions \cite{Liu2020}.

In global health emergencies, networks detailing interpersonal contacts have been pivotal in predicting disease transmission. The COVID-19 pandemic spurred the creation of composite models that integrate contact information with individual patient attributes \cite{hiram2022disease}. Within such models, nodes signify individuals, while links -- static or temporal/dynamic -- depict inter-individual interactions. Distinct individual features, such as health status (e.g., healthy or recovered), are encapsulated as node-associated feature vectors. Grounded in theoretical foundations of susceptible-infectious-recovered models \cite{hiram2022disease}, these approaches are nuanced and can account for real-world contact patterns. They allow for simulation and evaluation of public health response strategies, from containment measures to vaccination campaigns \cite{alguliyev2021graph,bryant2020modelling,stegehuis2016epidemic}. For example, designing a vaccination strategy targeting individuals based on contact behaviors could preempt outbreaks. {Since the design of a tailored vaccination strategy may save lives and control the epidemic spreading, we believe that more work should be done to improve these models by designing novel simulation algorithms which require less computational power. Actually, many simulation models require the inspection of all the nodes and edges for each simulation run,  making them difficult to run on very large graphs} \cite{Fortunato2010,hiram2022disease}.

\xhdr{Open questions for network-based precision medicine}
Despite notable advancements in network methods for precision medicine, several challenges remain. These include model benchmarking and comparison, integration of multimodal data from individual patients, and strategies to achieve the intricate equilibrium between preserving patient confidentiality and maximizing the utility of these approaches. Evaluating new methods is complex because establishing ground-truth, i.e., gold-standard or ``correct'', benchmarks against which various network strategies can be compared \cite{guo2021challenges} remains challenging. Evaluating precision therapeutics \emph{in vivo} presents even greater challenges, given the impossibility of retroactively altering treatment modalities for the same individual at a specific temporal junction. Garnering multimodal data about a single patient presents its own difficulties, as diverse data types vary in quality and completeness~\cite{wang2014similarity,zitnik2019machine}. In light of these complexities, there is a need for graph learning algorithms tailored for data-intensive multimodal networks. Importantly, new network embedding methodologies may provide simplification of these complexities into new modeling paradigms that are easier to comprehend and compute on \cite{Doria23,Doria23A,Xenos21}. Furthermore, it is imperative to foster computational paradigms adept at handling patient data in a manner that safeguards privacy while not compromising on scientific robustness and safety \cite{hunter2012reporting}.

Precision medicine stands poised to enable transformative shifts in disease diagnosis, therapeutic interventions, and overall patient care. Network methods and multimodal data integration are instrumental to these ambitions. Addressing intrinsic challenges related to small-sample datasets that lack statistical power
and magnifying methods' susceptibility to misinterpretation and unstable performance is paramount for furthering its nascent triumphs. Surmounting these obstacles requires interdisciplinary research involving network biology scientists, clinicians, and healthcare policymakers to ensure that precision medicine evolves as a paradigm for disease diagnosis, prevention, and treatment that works equally well for all patients by taking into account individual differences in lifestyle, socioeconomic factors, environment, and biological characteristics~\cite{all2019all}.

\vspace{-0.4cm}

%% file: 070.discussion_rtf.tex
Even the well-established network biology research topics/problems, such as network inference (Section \ref{sect:network_inference_comparison}), have many known limitations and thus open questions associated with them. The emerging research problems, such as network-of-networks analysis (Section \ref{sect:multimodal_networks_integration}) or determining how the explosion of large language models can benefit network biology, will have even more challenges associated with them, as expected, given that these problems have started to receive attention only recently; such challenges are discussed below. The emerging problems also bring exciting new opportunities. In the following sections, we build upon the discussion about limitations and open questions from the previous sections, link together common themes from the earlier sections, and complement the previous sections by introducing additional open problems and opportunities. 

\subsection{On methodological paradigms and empirical evaluation}

\xhdr{The need to compare different categories of approaches designed for the same purpose}
For several topics discussed thus far, a common theme has been that it remains unclear how specific categories of approaches for a given purpose compare to each other in terms of methodological (dis)advantages, as well as in which network analysis tasks or {biological/biomedical} applications they might be (in)appropriate to use. For example, with network alignment, methods from biological and other (e.g., social) network domains are rarely evaluated against each other (as discussed more below); with network-of-networks analysis, the existing approaches were proposed for different network analysis tasks or {biological/biomedical} applications and have not yet been compared to each other (Section \ref{sect:multimodal_networks_integration}); with hypergraph versus pairwise graph analyses, it remains unclear to what extent different tasks actually benefit from hypergraph-based methods (Section \ref{sect:higher_order_network_analysis}). 

Focusing more on network alignment, methods for this purpose introduced for biological networks have typically been thoroughly compared to each other (Section \ref{sect:multimodal_networks_integration}), including fair comparison of different approach categories, such as global versus local network alignment \cite{Guzzi2017,Meng2016}, pairwise versus multiple network alignment \cite{PNAMNA}, or alignment of static versus dynamic networks \cite{DynaMAGNA++}. On the other hand, network alignment methods introduced in network biology have rarely been compared to those introduced in other domains such as social networks, and vice versa, despite having similar if not the same goals -- mapping related nodes or network regions across compared networks. This could be because biological networks have significantly fewer nodes and are likely noisier than other (e.g., social) networks~\cite{eyuboglu2022mutual}. This could also be because networks in different domains contain different types of data, which makes the methods customized to their specific data types, rendering their comparison challenging or requiring methodological extensions and new developments. Or, it could be because developers of methods in different domains are from different scientific communities and may thus be unaware of each others' scientific discoveries (Section \ref{sect:additional_discussion}). In either case, it is critical to understand the methodological (dis)advantages of approaches from different domains. Their comprehensive and fair comparison could be a step in this direction, guiding the development of more powerful and possibly more generalizable network alignment approaches.

Network biology has traditionally relied on approaches that work directly on graph topology. In contrast, in recent years, the field has seen an increasing interest in network embedding -- be it via earlier spectral-based or diffusion/propagation/random-walk based methods or more recent deep learning methods -- which first transform graph topology into compact numerical representation vectors, i.e., embeddings, and then work on these graph representations (Section \ref{sect:networkml-intro}). A comparative study of non-embedding approaches that work directly on graph topology against network embedding methods was performed in a broad set of contexts: network alignment, graph clustering (i.e., community detection), protein function prediction, network de-noising, and pharmacogenomics \cite{Nelson2019embed}. The finding was that in terms of accuracy, depending on the context and evaluation measures used, sometimes direct, graph-based methods outperformed network embedding ones and other times, results were reversed; regarding computational complexity/running time, embedding methods outperformed direct, graph-based methods most of the time \cite{Nelson2019embed}. These indicate the need for a deeper combination of these approaches.

Also, network biology has traditionally relied on combinatorial or graph-theoretic techniques, i.e., on manually engineered or user-predefined topological features of nodes or graphs (the field has also relied on additional method types, e.g., those from the physics community within the field of network science, but these are not the focus of discussion here). For example, a prominent research problem of the graph-theoretic type that has revolutionized the field of network biology is counting graphlets/subgraphs in a graph;
various node-, edge-, or network-level features based on these counts are then applicable to many downstream computational tasks and {biological/biomedical} applications, as discussed in Section \ref{sect:higher_order_network_analysis}. More recently, network biology has benefited from the boom in deep learning (e.g., GNNs), which can automatically generate relevant network topological features prominently via graph representation learning (Section \ref{sect:networkml-intro}). It remains unclear which of graph-theoretic versus deep learning approaches (i.e., manually engineered versus automatically generated network topological features) are better and in which contexts. In other words, both approach categories seem to have merits depending on the context. Again, the question is how to combine them for improved performance.

As an example, graphlet-based and GNN-based analyses of protein structure networks were shown to outperform traditional non-network-based analyses of protein sequences and 3D structures in the tasks of protein structure comparison/classification and protein function prediction, respectively  \cite{GRAFENE,gligorijevic2021structure,NETPCLASS}. Only recently, the graphlet and GNN approaches were evaluated against each other when comparing protein structures, by the authors who proposed using GNNs for studying 3D structures \cite{gligorijevic2021structure}. They found that graphlet-based analyses greatly outperformed GNN-based analyses in accuracy, although they found the latter to scale better to denser protein structure networks \cite{berenburgyoutube1}. 

The relatively inferior performance of GNNs compared to graphlet-based approaches in that particular network-based protein structure comparison \cite{berenburgyoutube1} can potentially be elucidated as follows. Given that network comparison represents an NP-hard undertaking, a viable computational strategy that balances feasibility and efficacy involves the comparison of network substructures. Graphlets, by design, embody such an approach. Early GNNs were initially not designed for modeling subgraphs. So, it might not be surprising that popular GNN architectures cannot count graphlets and subgraphs and thus might not be the right methodological choice for specific scientific problems \cite{chen2020can}. Nevertheless, recent advancements in the field have yielded a spectrum of novel GNN methodologies tailored to subgraph modeling and enumeration. Theoretical underpinnings have emerged that show the expressive capacity of GNNs, delineating which classes of GNN architectures are proficient or deficient in quantifying specific subgraph structures \cite{bouritsas2022improving,chen2020can,tahmasebi2020counting,tahmasebi2023power,yu2023learning}.
For example, while message-passing GNNs have been popular architectures for learning on graphs, recent research has revealed important shortcomings in their expressive power. In response, higher-order GNNs have been developed that substantially increase the expressive power, although at a high computational cost \cite{tahmasebi2020counting}. These techniques demonstrate the potential to enumerate subgraphs, thus circumventing the established limitations of low-order (message-passing) GNNs while exploiting sparsity to reduce the computational complexity relative to higher-order GNNs \cite{tahmasebi2020counting}. Further, recent recursive pooling methods centered on local neighborhoods and dynamically rewired message-passing techniques~\cite{gutteridge2023drew} improve performance for tasks relying on long-range interactions. Finally, innovative methods based on graph transformers~\cite{ying2021transformers,zhang2022hierarchical} afford a spectrum of trade-offs between expressive capability and efficiency of machine learning models. 

Related to the above discussion, recent developments have highlighted the emergence of state-of-the-art geometric deep learning models trained on protein 3D structures~\cite{baek2021accurate,abramson2024accurate}. Many models focus on proteins' structural surfaces and some explicitly incorporate the underlying protein sequence or structural fold information~\cite{zhang2023full,dauparas2022robust}. Notably, these models have enhanced performance in various tasks associated with predicting interactions between proteins and other biomolecules~\cite{gainza2020deciphering,gainza2023novo,baek2024accurate}. These tasks encompass critical areas such as protein pocket-ligand prediction, prediction of PPI residues, ultrafast scanning of protein surfaces to forecast protein complexes, and the design of novel protein binders \cite{gainza2020deciphering, gainza2023novo}. Geometric deep learning methods that model protein 3D structures {\emph{as networks}} are promising. Such approaches were shown to outperform existing scientific methods traditionally used in a variety of tasks related to structure-based modeling and prediction of protein properties; {the existing methods included network approaches that are not} based on geometric deep learning \cite{stark2022equibind, wang2022learning, zhang2023protein}. The tasks in question included drug binding, PPI prediction, and protein fold, function, or reaction prediction/classification \cite{stark2022equibind, wang2022learning, zhang2023protein}.  

A potential avenue to handling different approach categories/paradigms, such as those discussed above, each with its own merits depending on the context, is to propose algorithmic improvements toward reconciling them. Another is to carry out empirical evaluation of different approaches in a variety of different contexts: at various levels of graph structure (e.g., node, edge, subgraph, or entire network), for diverse types of graphs (e.g., heterogeneous, dynamic, spatial), in different computational tasks (e.g., node classification, graph classification, link prediction), and different {biological/biomedical} applications (e.g., protein function prediction, cancer, aging, drug repurposing). The following sections discuss these two avenues in more detail.

\xhdr{Algorithmic improvements towards reconciling diverse methodological paradigms}
An algorithmic solution to handling different approach categories for the same purpose is to design hybrid methods that employ techniques from all associated disciplines. For example, deep learning methods can be combined with a network propagation approach to improve the embedding of multiple networks \cite{nasser2023bertwalk}. Alternatively, a theory that would unify different approach categories could be proposed. For instance, the field of neural algorithmic reasoning focuses on developing deep learning models that emulate combinatorial algorithms \cite{velivckovic2021neural}. {As a case in point, a transformer neural architecture, which was initially devised for natural language processing, has been repurposed to tackle the combinatorial traveling salesperson network problems \cite{bresson2021transformer} and graph-structured datasets~\cite{yun2019graph}.} A primary objective of this discipline is to investigate the capacity of (graph) neural networks to learn novel combinatorial algorithms, particularly for NP-hard challenges that necessitate heuristic approaches. Put differently, the aim is to ascertain if deep learning can extract heuristics from data more effectively, potentially superseding human-crafted heuristic methods that could demand years of dedicated research to formulate for NP-hard problems \cite{bresson2021transformer}. 

Another potential solution on the methodological level relies on the fact that current GNN approaches mainly adopt deep learning from other domains outside of network biology. As such, it is necessary to understand the correct inductive biases within a deep learning model that are representative of a biological mechanism under consideration. For example, can and should the hierarchical structures of ontologies, such as the GO or Disease Ontology, be incorporated into the GNN structure used for predicting proteins' functions or disease associations, respectively? Existing work on visible neural networks shows that such an attempt to incorporate a cell's hierarchical structure and function into the architecture of the deep learning model is effective and facilitates interpretability as the model's components naturally correspond to biological entities \cite{Thomas20,ma2018using}.  Even the hierarchical network-of-networks idea is not only useful as a potent new way to represent and analyze multiscale biological data as discussed in Section \ref{sect:multimodal_networks_integration}, but also as a novel graph representation learning methodology for popular network analysis tasks that are not necessarily of the multiscale nature. For example, there exist studies that take multiple networks as input, all at the same scale, and then perform the well-established tasks of graph embedding \cite{du2019mrmine} or classification \cite{wang2022imbalanced} via novel hierarchical approaches, e.g., a graph-of-graphs neural network \cite{wang2022imbalanced}, or matrix-factorization based data fusion \cite{iCell2019}. 

Another relevant question is how generalizable versus specific an approach should be. One frequent issue is selecting a suitable similarity measure. For instance, this issue arises when deciding which property of a graph should indicate the proximity of its nodes in an embedding produced by a GNN, or when discerning relationships between biomolecules for inferring correlation or regulatory networks by linking nodes with edges. Selecting an optimal similarity measure for a specific task or application often requires extensive empirical assessment, evaluating multiple measures against one another. It remains a challenge to discern whether a universal, principled similarity measure exists. The answer could potentially be specific to individual tasks or applications or broad categories of analogous tasks. The emphasis on generalizability also begs the question of its desirability; sometimes, the focus should be finely tuned to the specific task, application, or audience~\cite{ektefaie2024evaluating}. Furthermore, in some contexts, dissimilarity (or distance) might be more pertinent than similarity. For example, proteins can have opposing effects on each other despite working on the same functional goal~\cite{weber2020recent,badia2023gene,szklarczyk2023string}. As another example, neighboring edges might mean different things, such as up- versus down-regulation of genes. An essential consideration is the selection of distances with theoretical underpinnings that facilitate efficient optimization~\cite{Cao2013}, including distances that provably uphold the triangle inequality~\cite{ding2006transitive} and distances specified on smooth manifolds that yield symmetric positive semi-definite distance matrices~\cite{wang2018network}. Moreover, in typically high-dimensional spaces, the compromises entailed when our chosen distances forsake theoretical properties can be significant, potentially distorting interpretations and downstream analyses~\cite{Beyer1999, Radovanovic2010}.

\xhdr{{Uncertainty quantitation and confidence estimation}}
Uncertainty quantification presents a unique set of challenges. The inherent structure and complexity of network datasets introduce nuances not observed in other data modalities. The primary challenge lies in distinguishing between aleatoric (data-related) and epistemic (model-related) uncertainties while effectively mitigating potential biases that can distort predictive {performance~\cite{hullermeier2021aleatoric,zhao2020uncertainty}.}
Aleatoric uncertainty, stemming from inherent biological variation and limitations of experimental technology, encompasses variability arising from naturally random effects and natural variation intrinsic to the data~\cite{hullermeier2021aleatoric}. For instance, in PPI networks, inherent biological variability can lead to uncertainties in node or edge properties. On the other hand, epistemic uncertainty is engendered by a lack of knowledge or limited modeling assumptions. This type of uncertainty is particularly pronounced in graph-based tasks due to the myriad ways graphs can be represented, processed, and interpreted. For instance, different choices in GNN model architectures or graph pooling strategies can introduce varying degrees of epistemic uncertainty~\cite{hullermeier2021aleatoric}. 
Effectively quantifying and addressing these uncertainties is paramount for ensuring reliable and robust findings, especially when making critical decisions based on such models.

\xhdr{Additional considerations for proper empirical method evaluation: benchmark data, performance measures, code and data sharing, best practices}
Establishing appropriate benchmark data (including ground-truth data for training and testing/evaluating a predictive model), evaluation measures, and benchmark frameworks is critical to allow for systematic, fair, and unbiased method comparison. {Valuable efforts already exist (Table \ref{tab:ml-resources}). Nonetheless,} notably, such frameworks must allow for continuous evaluation as new methods and algorithms will continue to appear. Best practices and guidelines on assessment in network biology are needed. 

Lessons learned from challenges in biomedicine such as Critical Assessment of protein Structure Prediction (CASP) \cite{kryshtafovych2023new,Kryshtafovych2021,Moult1995}, Dialogue on Reverse Engineering Assessment and Methods (DREAM) \cite{meyer2021advances,saez2016crowdsourcing,stolovitzky2007dialogue}, and Critical Assessment of protein Function Annotation (CAFA) \cite{Jiang2016, Radivojac2013,Zhou2019} can perhaps help guide the development of best evaluation practices specific to network biology. Such challenges are a paradigm for unbiased and robust evaluation of algorithms for analysis of biological and biomedical data, which crowdsources data analysis to large communities of expert volunteers \cite{Costello2013,saez2016crowdsourcing}. Challenges are done in the form of collaborative scientific competitions. Through these, rigorous validation and reproducibility of methods are promoted, open innovation is encouraged, collaborative communities are fostered to solve diverse and critical biomedical problems and accelerate scientific discovery, the creation and dissemination of well-curated data repositories are enabled, and the integration of predictions from different methods submitted by challenge participants provides a robust solution that often outperforms the best individual solution \cite{saez2016crowdsourcing}. 

CASP is the earliest formal method assessment initiative in computational biology \cite{Moult1995}. While network biology approaches can be used for CASP's protein structure prediction and CAFA's protein function prediction problems, DREAM was explicitly initiated in response to a network biology need -- to reverse-engineer biological networks from high-throughput data \cite{stolovitzky2007dialogue}. Since then, numerous DREAM Challenges have been conducted spanning a variety of additional computational (not necessarily network) biology topics, including TF binding, gene regulation, signaling networks, dynamical network models, disease module identification, scRNA-seq and scATAC-seq data analysis, single-cell transcriptomics, and drug combinations\footnote{\url{https://dreamchallenges.org/}} \cite{meyer2021advances}. Note that in addition to these initiatives focused solely on computational biology tasks, there exist community benchmark frameworks for general graph-based machine learning that also handle some computational biology tasks, which could thus also serve as significant assets. An example is Open Graph Benchmark \cite{hu2021ogb,hu2020open} (Section \ref{sect:networkml-intro}), which includes the task of predicting protein function from PPI network data with fully reproducible results and directly comparable approaches using the same datasets\footnote{\url{https://ogb.stanford.edu/docs/leader\_nodeprop/\#ogbn-proteins}}. {Other examples are shown in Table \ref{tab:ml-resources}.}

Interestingly, some of the common themes that emerged from the original 2006 DREAM initiative \cite{stolovitzky2007dialogue} still hold to this date. The current biological network data may not be mechanistically accurate, yet they can still help understand cellular functioning. Exploring condition-specific biological networks is important because network properties can differ in different conditions. While there exist some highly trusted biological data (e.g., the reference HURI PPI network for humans \cite{luck2020reference})  that may serve as  ground truth for understanding (dis)advantages of network algorithms, synthetic network data that are much easier to generate will continue to be necessary for evaluating algorithm performance. However, experimentalists are unlikely to trust any scientific findings from synthetic data or computational approaches evaluated only on such data. Further, regarding ground-truth data for training and testing/evaluating a predictive model, it is critical to have available knowledge on both positive and negative instances in ground-truth data. Examples of the latter are PPIs or protein-functional associations that do not exist in cells. However, such negative instance data are hard to obtain in biology. 

To add to the discussion about ground-truth data, using the aging process as an example, ground-truth data about human aging have been obtained in one of two ways: via sequence-based homology from model species \cite{Magalhaes2009a} or via differential gene expression analyses in humans \cite{berchtold2008gene,jia2018analysis}.  In a recent study \cite{li2021improved}, only 17 genes were shared between the 185 sequence-based and 347 expression-based human aging-related genes. This poses several questions. How do we resolve such discrepancies with datasets on the same biological process resulting from different modalities/technologies, which likely exist in other applications as well? Given their high complementarity, perhaps integrating the different data types could yield more comprehensive insights into the biological process under consideration. However, if any of the other datasets are noisy, or if the different data types have different ``signatures'' (i.e., features) in a biological network, their integration could decrease the chances of detecting meaningful biological signals from the network compared to analyzing the different data types individually. Moreover, because different types of biological data collected via biotechnologies (e.g., genomic sequence data versus transcriptomic gene expression data versus interactomic PPI data) are likely to capture complementary functional slices of the given biological process, is it appropriate to use some of these datasets as the ground-truth data to validate predictions obtained via computational analyses of the other datasets? In our example of the aging process, is it appropriate to use sequence-based or expression-based aging-related knowledge to validate network-based aging-related gene predictions? Is this appropriate, especially because sequence-based and expression-based ``knowledge'' are also computational predictions, i.e., the result of sequence alignment and differential gene expression analysis, respectively? Also, is this appropriate because sequence-based knowledge about human aging are sequence orthologs of aging-related genes in model species? So, would any aspects of the aging process that are unique to humans be missed by the knowledge originally collected in the model species?

Another challenge with empirical evaluation is accurately estimating the absolute and relative performance of machine learning models and quantifying the uncertainty of performance estimates. Network data is inherently relational, thus inevitably violating the assumptions of independent and identically distributed data \cite{Neville2009, Neville2012}. Even further, the problems with long-tailed degree distribution in biological networks and homology between nodes require careful selection of training and test data when evaluating performance accuracy \cite{Hamp2015,Lugo-Martinez2021,Park2012}. 

Also, to allow for proper method evaluation, the authors of original methods must publicly release complete and easy-to-use code and data from their papers to allow for reproducing the initial studies and applying and evaluating a given method on new data \cite{heil2021reproducibility}. Journals and other publication venues should and typically do establish requirements for data and code sharing. Consequently, scientific communities have shown remarkable improvements regarding releasing open-source software and data. Yet, ensuring compliance remains an issue. For example, while  code or data might be released, they are sometimes incomplete or not easy to use. Or, there are instances when there might be a link (e.g., to  GitHub) provided in the corresponding publication to meet the publication venue requirements, but the link might point to a page that says ``under construction'', to an empty directory, or to a directory containing some files but without a transparent readme file on how to use the information provided. Who should ensure compliance with publication venue requirements, i.e., that complete and easy-to-use code and data are provided to ensure easy reproducibility? The editors of a venue publishing a given paper? The reviewers already volunteering their virtually non-existent ``free'' time to evaluate the paper's scientific merits for publication should thus probably not be expected to invest even more effort to verify that the code and data can be run correctly. The authors? The future readers of the paper who might be interested in using the method? If the latter two, what should be the repercussions if it is found that the code or data do not exist or are not possible or easy to use? On a related note, how long after publication should the authors be required to maintain the project code and data and respond to related email inquiries? Hosting of the code and data is not an issue for authors due to availability of archival data repositories such as Zenodo. However, actively maintaining the code and data is an issue, and this is directly related to whether and how long after the project completion the funding by the federal agencies and others might be available for this purpose.

{Complete transparency in all decisions (from graph construction to analysis) is crucial.  Workflow management systems, such as Nextflow~\cite{di2017nextflow} and Snakemake~\cite{koster2012snakemake}, can enable rapid prototyping and deployment of computational workflows by combining software packages and various tools. Clear documentation, open-source sharing of code and algorithms, and making raw and processed data available can ensure that results are not just a one-off finding but can be consistently reproduced and built upon by the broader scientific community.} 

\subsection{On missing data}

\xhdr{Network completeness and interaction causality}
Much of network biology relies on aging technologies with notable limitations. Focusing on physical PPIs, biotechnologies such as yeast two-hybrid systems \cite{Fields1989}, cross-linking mass-spectrometry \cite{Piersimoni2022}, and structural determination of protein complexes \cite{Jacobsen2007,Rhodes1993,Saibil2022} have collectively generated systems-level data that have led to critical methodological advances in network biology. Of course, these efforts to obtain the physical interactome have been complemented by valuable data collection and network inference efforts related to systems-level correlation networks. However, as computational methods are now maturing, the data are starting to lag. High-resolution, high-throughput data-generating technologies, capable of directly identifying pathways and order of molecular events in various experimental and clinical contexts, are the next frontier for deeper understanding of molecular systems.

There is a need to expand from physical and correlation networks toward causal relationships \cite{belyaeva2021causal} or simulatable kinetic models \cite{karr2012whole}. For this, biotechnologies for data collection need to be improved to allow for higher-quality data to build better causal networks and more complete networks. This will also require the development of new (categories of) approaches that can handle the captured causality. Even if/when we have high-quality causal networks and efficient and accurate methods for their analysis, will this suffice to understand biochemical mechanisms? When one knows biochemical mechanisms, one can infer causality. However, causality {might not necessarily allow for fully understanding} biochemical mechanisms.

\xhdr{Algorithmic research to guide data generation efforts}
It will likely be beneficial to integrate multi-omic network data with BKGs to offer precise and targeted treatments for rare diseases \cite{alsentzer2022deep}. Such network data with richer semantics will more directly help suggest biological hypotheses \cite{sanghvi2013accelerated,wang2023scientific} or support iterative data generation and analyses through active learning \cite{sverchkov2017review,zhang2022active}. Informing laboratory experiments using predictions from computational studies could be a path forward to build more complete and accurate data, which could lead to developing new, more advanced network analysis methods to further inform and improve laboratory experiments. 

How network biology (primarily algorithmic research) can best support the collection and analysis of multimodal data is quite an important question, especially when collecting multimodal data for the same individuals, including building personalized  (i.e., individual-specific) networks. An answer here could be to first figure out what question will be asked in which task/application and then design a data collection strategy. One might want to define optimal datasets. Or, one might want to find unifying factors within data modalities; this is precisely why there is a need for multimodal data for the same individuals, at least some of the data/individuals. This might require systematic, comprehensive, and well-funded consortia efforts. Perhaps algorithmic approaches such as active learning can help prioritize what data should be collected, e.g., from specific populations or about particular biological {functions. As} success in experimentally collecting or computationally inferring various types of biological networks continues to improve, research efforts likely should shift towards obtaining a predictive understanding of personalized networks. Moreover, even within a single individual, molecular networks vary across tissues and cell types, posing additional challenges in defining an individual-specific network. 

\xhdr{Network dynamics}
Another data component that is currently missing or is very scarce is network dynamics. Various types of time-dependent perturbation data could help infer dynamic biological networks. {Examples of tasks/applications  that have benefited from dynamic network analysis in biology are as follows.} 

One example is the task of network alignment: unlike traditional network alignment that has compared static networks (Section \ref{sect:multimodal_networks_integration}), recently, the problem of aligning dynamic networks has been defined, and several algorithms have been proposed for solving the newly defined problem \cite{GoTWAVE,DynaMAGNA++,DynaWAVE}.  The challenge here is the lack of experimentally obtained dynamic biological network data, which is why such methods have been evaluated on synthetic networks, computationally inferred dynamic biological networks, or dynamic networks from other domains \cite{GoTWAVE,DynaMAGNA++,DynaWAVE}. 

Another example is a recent network-based study of the dynamics of the protein folding process \cite{newaz2022multi}. A key challenge is the lack of large-scale data on protein folding intermediates, i.e., 3D conformations of a protein as it undergoes folding to attain its native structure.  Experimental data of this type are lacking even on the small scale \cite{newaz2022multi}. Traditional computational, simulation-based studies, as well as the recent network-based effort \cite{newaz2022multi}, all approximate the folding intermediates of a protein from the protein's final (or native) 3D structure. Obtaining the actual protein folding intermediates experimentally is unlikely to happen any time soon, especially at a large scale, so computational efforts will be needed. With recent breakthroughs in protein structure prediction, e.g., AlphaFold \cite{jumper2021highly}, this need represents an excellent opportunity for computational research to help obtain, model, and analyze the resulting dynamic data.  

{A further} example is a dynamic network analysis of the aging process, i.e., predicting new aging-related genes from a dynamic aging-specific PPI network (Section \ref{sect:network_inference_comparison}). Here, a key challenge is that shockingly, using {\emph{newer}} aging-related gene expression and PPI network data obtained via newer {and thus higher-quality} biotechnologies to infer a dynamic aging-specific network does not yield more accurate aging-related gene predictions than using {\emph{older}} data {of the same type} from over a decade ago when dynamic network analyses of aging were pioneered \cite{li2022towards}. It was also observed in a different study on active module identification that using newer network data typically did not lead to more biologically meaningful results \cite{lazareva2021limits}. Going back to aging, it remains unclear whether the issue is with gene expression data, PPI network data,  methods for integrating the two to computationally infer a dynamic aging-specific network, network methods used for feature extraction from the aging-specific network, ground-truth data on which genes are aging- versus non-aging-related, or something else entirely \cite{li2022towards}. 

{As our final example, we discuss quantitative and qualitative mathematical modeling of network dynamics from the systems biology perspective \cite{le2015quantitative, kestler2008network}. Quantitative formalisms provide a precise description of the evolution of the system, including its temporal aspects; they are strongly dependant on the availability and precision of the required parameters.  At the other end of the spectrum, qualitative (logic) frameworks have the advantage to be simpler, with no requirement for quantitative parameters, allowing analytical analyses. Logical models allow coarse-grained descriptions of the properties of the biological network and bring out key actors and mechanisms controlling the dynamics of the system \cite{maheshwari2017framework}. Recent efforts use -omics data, including single-cell transcriptomes, to construct or contextualize Boolean models \cite{herault2023novel,montagud2022patient,schwab2021reconstructing}.}

\xhdr{Towards inclusive and equitable precision medicine}
Progress in computational (including network) biology and biomedicine has been hindered by a lack of -omics data encompassing vast human diversity \cite{cruz2023importance}. Underrepresentation of human genetic diversity has drastically weakened the biological discoveries that would benefit all populations, leading to health disparities. The traditional one-size-fits-all healthcare model meant for a ``typical'' patient may not work well for everyone. In response, the National Institutes of Health has aimed to invite one million people across the United States to help build one of the most diverse health databases in history, welcoming participants from all backgrounds through the ``All of Us'' program\footnote{\url{https://allofus.nih.gov/}}. Inclusivity is at the core of the program: participants are diverse in terms of their races, ethnicities, age groups, regions of the country, gender identity, sexual orientation, socioeconomic status, education, disability, and health status. The data collected through the program is expected to lead to discoveries on how our biology, environment, and lifestyle affect our health. Unlike traditional research that has focused on a particular disease or group of people, this program aims to build a diverse database that can inform thousands of studies on a variety of health conditions. Availability of inclusive and diverse -omics data, design of research studies that intentionally and carefully account for such data, and development of computational methods and evaluation frameworks that handle such data in a fair and unbiased manner will be critical for advancing computational biology and biomedicine for all populations and reaching health equity.

Beyond the issue of underrepresentation, certain populations are intrinsically limited in size, such as rare diseases, which are inherently limited in clinical cases~\cite{banerjee2023machine}. Studying a substantial fraction of a small population may still result in data that do not yield health outcomes comparable to those from larger populations. In such scenarios, amassing more data may not be feasible, leading to small-sample datasets that can lack statistical power and magnify the susceptibility of computational models to misinterpretation and unstable performance. Network analysis techniques can play a pivotal role in addressing this challenge. Techniques such as few-shot machine learning~\cite{alsentzer2022deep} and domain adaptation~\cite{he2023domain} for network methods are instrumental in enabling computational models to learn patterns from small datasets and generalize to newly acquired data. Such models can adapt and generalize across diverse populations, thereby enhancing the robustness and applicability of health outcomes derived from datasets with small numbers of samples. 

\subsection{Other major future research advancements}

\xhdr{The interface between network biology and large language models}
{Large language models (LLMs), such as ChatGPT and GPT-4, create opportunities to unify natural language processing and knowledge graph reasoning~\cite{fatemi2023talk,pan2024unifying}, owing to their wide-ranging applicability.} Nevertheless, LLMs often serve as black-box models, presenting limitations in comprehensively capturing and accessing factual knowledge. In contrast, BKGs are structured knowledge models that systematically store extensive factual information. BKGs have the potential to enhance LLMs by providing external knowledge that aids in inference and bolstering interpretability. However, constructing BKGs is intricate and dynamic, posing challenges to existing methods in generating novel facts and representing previously unseen knowledge. Thus, an approach integrating LLMs and BKGs could emerge as a valuable strategy, harnessing their strengths in tandem~\cite{pan2024unifying}.

The potential synergies between traditional text and structured knowledge graphs are becoming increasingly evident. Language model pretraining has proven invaluable in extracting knowledge from text corpora to bolster various downstream tasks. Yet, these models predominantly focus on single documents, often overlooking inter-document dependencies or broader knowledge scopes. Recent advances~\cite{mcdermott2023structure,yasunaga2022linkbert} address this limitation by conceptualizing text corpora as interconnected document graphs. By placing linked documents in shared contexts and adopting self-supervised objectives combining masked language modeling and document relation prediction, such methods can achieve considerable progress in tasks like multi-hop reasoning and few-shot question answering. On a parallel front, while text-based language models have garnered substantial attention, knowledge graphs can complement text data, offering structured background knowledge that provides a useful scaffold for reasoning. In an emerging line of inquiry, studies~\cite{yasunaga2022deep} explore self-supervised paradigms to construct a unified foundation model, intertwining text and knowledge graphs. These approaches pretrain models by unifying two self-supervised reasoning tasks, masked language modeling, and link prediction, marking an exciting direction for future advancements in network biology.

LLMs, traditionally associated with the processing of natural language, possess a flexibility that extends their utility beyond text data~\cite{luo2022biogpt}. The underlying architectures, especially transformer-based designs like BERT and GPT variants, can be adapted to learn from any sequential data. In biology, this adaptability implies that LLMs can be trained on biological sequences, such as DNA, RNA, and proteins~\cite{lin2023evolutionary,rao2019evaluating,xu2022peer}. Rather than processing words or sentences, these models can assimilate nucleotide or amino acid sequences, thereby capturing intricate patterns and dependencies in genomic and proteomic data~\cite{dauparas2022robust,lin2023evolutionary,mcdermott2023structure,meier2021language}. These cross-disciplinary advances in LLMs highlight their potential to advance the frontiers of computational biology. In addition to large sequence-based pretrained models like LLMs, an emerging area of structure-based pretrained models is concerned with generating new network structures, such as protein and small molecule networks~\cite{bennett2023improving,gainza2023novo,rodrigues2022csm,townshend2021geometric,wang2022scaffolding}.

\xhdr{{Interpretabilty}}
{Interpretability in network biology involves elucidating mechanisms of disease and health, such as tumor growth and immune responses. However, deep graph learning models are black-box systems with limited immediate interpretability as they produce outputs through a series of complex, non-linear transformations of input data points. This poses challenges in domains where clear insights are imperative. For instance, while dimensionality reduction techniques and graph representation learning algorithms produce compact latent feature representations of high-dimensional data and graphs, they often sacrifice the interpretability of the features they produce. Conversely, graph-theoretic signatures, which capture network motifs, graphlets, or other substructures, can amplify understanding of networks by identifying relevant structural patterns.}  

Future research directions in interpretability must focus on integrating domain-specific knowledge into model training and evaluation. By directly incorporating biological constraints and prior knowledge into model architectures, we can enhance interpretability without compromising predictive performance. Additionally, developing explainable techniques tailored explicitly for network biology is crucial. Exploring hybrid models combining interpretable statistical models with deep learning approaches is another promising avenue. Such models can leverage the strengths of both types to produce interpretable and accurate predictions. Likewise, creating advanced visualization tools that effectively convey complex model outputs and biological insights to researchers and clinicians is essential. These tools should be intuitive and enable interactive exploration of model predictions and features.

\xhdr{{Reproducibility}}
Reproducibility in network biology research is a multifaceted challenge due to several reasons. (1) Graph construction: How a graph is constructed can drastically impact the insights drawn from it. For example, consider the problem of inferring an association PPI network. The decision to include only direct interactions versus both direct and indirect interactions can lead to vastly different network topologies. Choosing a threshold to determine an edge (e.g., a particular strength of interaction or confidence level) can also significantly alter the graph. (2) Edge definitions: What constitutes an edge can be subjective and is often based on the specific context. In a gene co-expression network, for instance, the definition of an edge might be based on a particular correlation coefficient threshold. A slight variation in this threshold can lead to including or excluding numerous interactions, thus changing the network's structure and potentially its inferred properties. (3) Latent embeddings: Graph-based machine learning methods used to compute embeddings can have a significant effect on the results. Different embedding techniques capture different types of structural and feature-based information, leading to variations in tasks like node classification or link prediction. (4) Dynamic nature of biological networks: Biological systems are inherently dynamic. A PPI network at one point in time or under one set of conditions might differ from the network under another state. Thus, reproducing results requires the same methodology and the same or equivalent biological conditions. (5) Finally, graph sampling: In many cases, a subgraph or sample is taken due to the massive size of networks or computational constraints. The method and randomness inherent in this sampling can lead to non-reproducible results if not carefully controlled.

\xhdr{Towards wide adoption and translation of algorithmic innovation into practical and societal impact}
The recommended method evaluation and data generation improvements discussed above are needed not just for method developers -- typically, computational scientists -- to be able to properly evaluate their new approaches against existing ones, but even more importantly, for adoption by end users -- experimental scientists and in the long run, clinicians, healthcare workers, and patients (Section \ref{sect:additional_discussion} comments more on this topic, including training needed for non-computational folks to use network approaches). The disconnect between computational and experimental scientists, even those dedicated to the common scientific goals \cite{Ramola2022}, suggests that efforts are necessary to overcome both technical and social challenges in interdisciplinary research fields. Computational scientists might need to consider not only traditionally algorithmic evaluation measures, such as precision,  recall, and other performance criteria, but also measures that evaluate the utility and feasibility of integrating methods into scientific and clinical workflows~\cite{huang2023zero,huang2022artificial}. Additionally, computational scientists are primarily incentivized to develop new algorithms and prototype software. In contrast, experimental and clinical scientists expect tools that are robust, trustworthy, and exhibit few glitches in practice. Authoritative evaluations, carried out by independent and interdisciplinary researchers on tasks directly relevant to downstream applications, are essential~\cite{choobdar2019assessment,marbach2012wisdom}. Rapid and broad dissemination of these evaluations, recommendations, and guidelines for best practices should be prioritized in network biology.

\xhdr{Major milestones in network biology} 
The pinnacle of success for network biology would likely be a comprehensive and dynamic understanding of the entire cellular or organismal interactome across different conditions and life stages. This would include PPIs, gene regulation, metabolic pathways, cell signaling, and more. We can imagine a complete map of every biological interaction in an organism, from the level of genes and molecules up to tissues and organs, with the ability to zoom in on details and see dynamic changes over time or under different conditions. Another significant milestone would be the seamless integration of network biology with other disciplines to provide a holistic understanding of life. This means connecting the molecular interactome with tissue-level networks, organ systems, and inter-organismal interactions, such as those seen in symbiosis or ecosystems. From a practical standpoint, a significant success measure would be the application of network biology insights to develop novel and more effective therapeutic interventions. This could mean identifying critical network nodes or interactions to target diseases, leading to innovative treatments.

Drawing parallels from the reference human genome, the equivalent for network biology could be a reference interactome---a standardized and comprehensive map of all known biological interactions within a human cell. This would serve as a baseline for studying disease, development, aging, and other biological processes. Any deviations from this reference in specific cell types, conditions, or diseases could be studied in detail.

Just as AlphaFold ~\cite{jumper2021highly} has made waves in predicting protein structures, a comparable success in network biology might be the development of tools that can accurately predict the emergent properties of a biological system from its underlying network. Given a set of interactions, this would mean the tool could foresee the system's response to a drug, its behavior under certain conditions, or its evolution over time.

%% file: 080.community_rtf.tex
The question of who are network biologists or computational biologists is hard. Ideally, a computational biologist would have the interest and knowledge to both develop core computational methods and understand fundamental biological mechanisms. That raises the question of how to properly train more of such researchers to advance computational biology, including its subarea of network biology that models and analyzes biological systems as networks. For example, based on the personal experience of some of the authors of this paper, in a network biology course, computationally-focused students might enjoy computational but not biological aspects  (e.g., in a general network science course, students typically choose a non-biology domain to work on, such as technological or social networks). In contrast, biology students might enjoy biological but not computational aspects. So, efforts might be needed to convince students to be genuinely excited about both developing computational approaches and understanding biological mechanisms. Systematically identifying and addressing gaps in current computational biology training programs or starting new interdisciplinary training programs might be needed, along with appropriate support and resources from funding agencies. 

Some of these gaps are as follows. An essential part of efficient training would be to have robust, well-known, and trustworthy software tools that are readily available and easy to use, especially by those who are not proficient in computing; clearly, both developing and sustaining such software requires resources. 
Similar holds for building and making available datasets easily accessible by people who are not proficient in biology to help them get involved easily. Another important part would be exposing students to interdisciplinary collaborative teams to train them to work together on the same research questions with scientists from different disciplines.

Another vital part of training relates to hiring and promoting computational biology faculty who would offer the training. A challenge here, based on the personal experience of some of the authors of this paper, seems to be as follows. When hiring a computational biologist in a traditional computationally-focused department (e.g., computer science, applied mathematics, statistics, or physics), someone who is more trained in biology may be viewed as not enough of a computational scientist, even when they are proficient in using existing computational methods to uncover new biological knowledge and possibly also at least occasionally develop new computational methods for studying biological systems. Similarly, in a traditional biology-focused department, a  more computationally trained person may be viewed as not enough of a biological scientist, even when they evaluate their new computational methods on biological data and possibly at least occasionally yield new knowledge about biological systems. Yet, both kinds of candidates can be great for both department types. Hence, hiring and promotion groups might need to think differently about interdisciplinary computational biology research. This is especially true in departments where these groups do not have computational biologists or where there are no specific, interdisciplinary departments like biomedical data science or computational biology.

{There exists an additional challenge} even when focusing on computationally-oriented researchers within computational biology. Scientific communities that could benefit (from) the field of network biology include graph theory, network science, data mining, machine learning, and artificial intelligence. These communities often use different terminology for the same concepts (e.g., network alignment versus graph matching or graph clustering versus network community detection). Distinct scientific communities may all analyze biological network data, or address identical computational challenges across various application domains, such as biological versus social networks. However, they often do not attend the same research forums. For instance, attendees of the prominent computational biology conference, Intelligent Systems for Molecular Biology (ISMB), might not necessarily participate in data mining conferences like Knowledge Discovery and Data Mining (KDD) or artificial intelligence conferences such as Neural Information Processing Systems (NeurIPS), and vice versa. Consequently, advancements in one domain might remain obscure in another. Organizing scientific symposia to convene computational scientists from traditionally distinct network biology communities, focusing on universally relevant topics, could help bridge this gap.

The above discussion items can be seen as diversity-focused, be it diversity in one's training and skills or scientific communities they belong to~\cite{nielsen2018making}. Many other aspects of diversity exist in science, and we focus on some of them here. The  International Society for Computational Biology (ISCB) is a globally recognized entity advocating for and advancing scholarship, research, training, outreach, and inclusive community building in computational biology and its professions. This is why we rely on ISCB's demographic statistics to represent the current state in the computational biology field. According to a demographic survey of the ISCB membership, whose results are publicly available in the 2022/2023 ISCB Equity, Diversity, and Inclusion (EDI) report\footnote{\url{https://www.iscb.org/edi-resources}}, among those who responded, 32.8\% indicated ``female'', 60\% indicated ``male'', 0.4\% indicated ``non-binary'', and 6.8\% indicated ``prefer not to declare''. Regarding ethnic origin, in the same report, 53\% of those who responded with anything but ``prefer not to declare'' indicated a non-European descent. Some additional EDI statistics are as follows. At the time of the 2020/2021 ISCB EDI report (the latest report that offered this type of information), 41\% of the ISCB Board of Directors were female, and 57\% of the Executive Committee (elected officers) were female; 61\% of selected keynote speakers at the Intelligent Systems for Molecular Biology (ISMB), ISCB's flagship and most prestigious conference, were female since 2016. Regarding ISCB awards, fellows election, and other honors, the final selection shows a good gender balance that reflects the membership. However, during the nomination stage, in 2022/2023, for the innovator award, senior scientist award, and fellows election, 22\%, 28\%, and 25\%  of the nominees were female, respectively, compared to 32.8\% of the entire ISCB membership being female. ISCB does not have such data yet on ethnicity.

Enhancing awareness and mitigating biases when nominating candidates for honors or inviting candidates as conference speakers is a pathway to improving diversity in the computational biology field. Another more ambitious goal is to achieve diversity statistics in the field that mirror those of the general population. This should be accomplished  for all of undergraduate students, graduate students, postdoctoral fellows, and faculty (across various ranks), not only by addressing the `leaky pipeline' issue~\cite{alper1993pipeline,sarraju2023leaky}, but also by identifying and eliminating institutional barriers to establish an inclusive support infrastructure~\cite{stevens2021fund}. This might only be achievable over a longer period. Also, biology-focused subfields of computational biology are currently more gender-diverse than its computationally-focused subfields. Thus, diversity in computational biology might be more readily achieved by recruiting trainees from biology-focused subfields and equipping them with the requisite computational skills rather than the reverse. However, sourcing from computational subfields remains essential. Yet, disciplines like computer science, mathematics, and physics can act as gatekeepers and entering these fields without the appropriate background can be challenging~\cite{mervis2022fix,torbey2020algebra}. Because innovative concepts can emerge from diverse sources and all individuals, it is imperative to eliminate gatekeeping barriers. 

Additional diversity-related challenges include the need to recognize and mitigate potential implicit biases; limited access to registration and travel funds to conferences based on their locations, especially for those in middle and low-income countries; current lack of ethnicity data to evaluate diversity efforts of computational biology conferences and communities, including ISCB; empirical research into equity in science, etc. Systematic and properly funded initiatives by universities and professional societies are necessary to achieve this. And so are individual efforts by the members of the scientific community. Everyone should be responsible for contributing to joint diversity efforts for the field to make significant and sufficient progress.